\documentclass[]{aa}  

\usepackage{txfonts}
\usepackage{graphicx}
\usepackage{color, colortbl}
\usepackage{subcaption}
\usepackage{caption}
\usepackage{amsmath}
\usepackage{hyperref}
\usepackage{siunitx}
\usepackage{pifont}
\usepackage{tabularx}
\usepackage[export]{adjustbox}
\usepackage{natbib}

\newcommand{\customcite}[1]{\citeauthor{#1} (\citeyear{#1})}

\begin{document}

   \title{Solar synthetic imaging: Introducing denoising diffusion probabilistic models on SDO/AIA data}

   \author{F. P. Ramunno
          \inst{1, 2}\thanks{Email address: francesco.ramunno@fhnw.ch}
          \and
          S. Hackstein\inst{1}
          \and
          V. Kinakh\inst{2}
          \and
          M. Drozdova\inst{2}
          \and
          G. Quétant\inst{2}
          \and
          A. Csillaghy\inst{1}
          \and
          S. Voloshynovskiy\inst{2}
          }

   \institute{Institute for Data Science, University of Applied Sciences North Western Switzerland (FHNW), 5210 Windisch, Switzerland
         \and
             Department of Computer Science, University of Geneva, 1211 Geneva, Switzerland
             }

   \date{Received September 1, 2023; Accepted March 3, 2024}

 
  \abstract{For the luck of humanity, there are way less big solar flares than small ones. Even if these are good news, this makes it challenging
   to train machine learning algorithms able to model solar activity. As a result, solar monitoring applications, including flare forecasting, suffer from this lack of input data.  To overcome this issue, generative deep learning models can be utilised to produce synthetic images representing solar activity and thus compensating the rarity of big events. This study aims to develop a method that can generate synthetic images of the Sun with the ability to include flare of a specific intensity. To achieve our goals, we introduce a Denoising Diffusion Probabilistic Model (DDPM). We train it with a carefully crafted  dataset from the AIA (Atmospheric Image Assembly) instrument on the SDO spacecraft, specifically the 171 Å band, which captures images of coronal loops, filaments, flares, and active regions.  GOES X-ray measurements are employed to classify each image based on the solar flare scale (A, B, C, M, X), after selecting the flaring images from  AIA using the Heliophysics Event Knowledgebase, which allows for temporal localisation of the flaring events. The generative model performance is evaluated using cluster metrics, Fréchet Inception Distance (FID), and the F1-score. We demonstrate state-of-the-art results in generating solar images and conduct two experiments that use the synthetic images. The first experiment trains a supervised classifier to identify those events. The second experiment trains a basic solar flare predictor. The experiments demonstrate the effectiveness of additional synthetic samples to addressing the problem of imbalanced datasets. We believe this is only the beginning of DDPM use with solar data. It remains to gain a better understanding of the generation capabilities of the denoising diffusion probabilistic models in the contest of solar flare predictions and apply them to other deep learning and physical tasks, such as AIA to HMI () image translation. }


   \maketitle

\section{Introduction}
\label{sec:introduction}

Solar flares pose a threat to Earth and its inhabitants due to their ability to induce geomagnetic storms that can disrupt modern technological infrastructure. Their effects can have significant consequences for various technologies, such as the communication systems, causing radio communication disruptions, especially at high frequencies. This can impact on airline communications and those of emergency services and others \citep{knipp2016, redmon2018september, xu2022characteristics}. Solar flares are also dangerous for astronaut safety \citep{Smith_2007, Fargion_2019} increasing their risk of radiation-related health issues and also for satellite operations \citep{gopalswamy2023solar} leading to temporary loss of service due to the increase in radiation. 

This implies a need to build forecasting and nowcasting algorithms for the prediction of their arrival and mitigation or nullification of their effects \citep{Cicogna_2021, guastavino2022operational, Huwyler_2022, tlatov2023groundbased}. However, as we know, algorithms are only as good as the data they rely on. 

A major problem in the prediction of solar flares begins with the fact that the intensity of solar flares is inversely proportional to their occurrence rate; indeed the most dangerous are the rarest events \citep{Aschwanden_2012}. This results in unbalanced datasets \citep{imbdataset}, which create significant challenges in effectively training an algorithm to predict these events. The lack of generalisation and strong bias towards more frequent flare classes can be attributed to the difficulty in obtaining datasets that equally represent the various classes. For a model to successfully grasp the required information and make accurate predictions, it is essential to have datasets that represent each class equally in order to avoid biases and improve generalisation. Furthermore, understanding flares is also of interest to studies of also interest particle acceleration, plasma ejection and their morphology in different wavelengths \citep{refId0, Collier_2023}. Thus, being able to study the highest energy flares with a large amount of data and with the ability to control all the characteristics of images of the Sun can also be useful in order to better understand the triggers that lead to these high energetic events and also their evolution.

Recently, there has been an increase in the popularity of generative models \citep{rombach2022highresolution, ramesh2022}. Consequently, it is interesting to explore the feasibility of training a model capable of recognising and generating the different patterns that define solar activities. Such a task holds the potential to change the utilisation of synthetic data, extending beyond just class representation, and facilitating the discovery of novel physics.

This work focuses on the development of a method that can generate synthetic images of the Sun, whilst allowing the user to control the presence of a flare of a given intensity. Our aim is to investigate the capacity of the model to distinguish the solar features that can potentially trigger such events and to be able to generate them. 

There have been various attempts to use deep learning generative models \citep{Liu_2022, deng2021, dash2021high}, mainly for image-to-image translation purposes \citep{Salvatelli_2022} (e.g. AIA (Atmospheric Image Assembly) to HMI (Helioseismic and Magnetic Imager)).  In recent years, Generative Adversarial Network (GAN) has been the state-of-the-art model for image generation and variations of this task (e.g. image-to-image translation and image in-painting \customcite{Chen2022}). Unfortunately, GANs present some limitations. The most important for our work is the fact that GANs suffer significantly from mode collapse. Thus, if some classes are under-represented, it is more likely that the model is going to ignore them with a preference for the most populated classes. This is why we turn our attention here to the denoising diffusion probabilistic models (DDPMs) \citep{ho2020}. \customcite{classifierguidance} analysed how diffusion models can overcome the GANs limitations. The latent space learned by a diffusion model has been shown to be useful in discriminative tasks such as classification and anomaly detection \citep{zimmermann2021scorebased, wolleb2022diffusion}. As a result, the image quality results obtained with diffusion models are better than those with GANs as shown in \customcite{rombach2022highresolution}. Furthermore, most importantly, the diffusion models are better in capturing the ground-truth distribution of the data analysed by metrics such as the FID (Fréchet inception distance \citep{FID}), which helps in cases where there are under-represented classes.

In the present work, we investigate the capabilities of the DDPMs, which have already proven to be valuable in other, diverse application domains such as computer science, medicine and astrophysics \citep{um2023dont, huy2023denoising, karchev2022stronglensing}. This method allows us to generate synthetic images of the Sun given a specific label from the GOES classification system .
The labels are used to guide the process during the sampling towards the generation of a specific image of the Sun with the correct amount of activity.

To the best of our knowledge, we are the first to introduce the concept of DDPMs in the field of heliophysics and the first to guide the sampling process being able to fill the unbalanced high energy solar flare classes (e.g., M- and X-flare class). 

We use images obtained by the Solar Dynamics Observatory (SDO) telescope in the training procedure. As a future work, with the results of this project, we aim to demonstrate the use of the synthetic images of a particular flare class to train machine learning algorithms for image classification and flare forecasting/nowcasting and to investigate these phenomena more extensively based on more available data.

This paper is organised as follows.
In Sect.~\ref{sec:data_source} we introduce the datasets used. In Sect.~\ref{sec:back_ddpm} we explain the DDPM together with the classifier free guidance technique. In Sect.~\ref{sec:experiments} we analyse our setup and our experiments; we discuss then their results them in Sect.~\ref{sec:results}. We present two different uses of the model in Sect.~\ref{sec:discussion} and finally conclude in Sect.~\ref{sec:conclusions}.
\section{Dataset}
\label{sec:data_source}

In this work, we use three datasets: (1) the version 2 of the SDO Machine Learning Dataset (SDOMLv2), which is an update of version 1 by \customcite{Galvez_2019}, available at a dedicated Github repository \footnote{\url{https://sdoml.github.io}}, and provides full Sun images; (2) the GOES X-ray sensor data, which we use to retrieve the X-ray emission; and (3) the Heliophysics Events Knowledgebase \citep[HEK]{Hurlburt2010}, which we use as event recording notifier.

\subsection{SDO machine learning dataset}

The origin of the data used is the AIA \citep{aia}, an instrument on board of the SDO satellite. AIA records full-disc images of the solar photosphere, chromosphere, and corona in two ultraviolet (UV) channels and seven extreme ultraviolet (EUV) channels.
However, the AIA data cannot be used directly for ML; first they need to be preprocessed to be spatially coregistered, to have equal angular resolutions, and to be corrected for instrumental effects. Therefore, a subset called SDOML \citep{Galvez_2019} has been created so that it can be directly used for machine learning studies. In this study, we are using SDOMLv2, which is updated to account for a change in calibration after 2019, uses of the new zarr format, and adds the data up to the present day. 

In this study, we are working with 64x64 images, because of the heavy computation of the model as explained in Sect. \ref{sec:back_ddpm}. We are conscious that the image size will need to be increased for an operational study. As we show in Section \ref{sec:results}, the 64x64 images are still able to model the solar activity; however, in a future study, we would like to explore also the impact of image size on the applicability of the synthetic images. We are using the AIA \SI{171}{\angstrom} channel. This band is chosen because a broad range of solar activity is visible there , with many features, and therefore it is interesting to test whether or not if the generative model is able to reproduce this activity due to the complicated nature of this channel. In addition, the \SI{171}{\angstrom} channel is also used to compare the results with the work by \customcite{marius}, who also used this channel for an anomaly-detection task based on a generative model. This allows us to determine the ability of the DDPMs to generate images of the Sun and to examine their quality.

\subsection{GOES X-Ray sensor}
Since 1986, a series of GOES spacecraft have been taking measurements of soft X-rays in two energy bands (X-Ray sensor A (XRSA) 0.5-\SI{4}{\angstrom} and XRSB 1-\SI{8}{\angstrom}). The XRSB channel is used to mointor the solar flares and to determine their magnitude. We downloaded the data from 2011 to 2019 with the Python library SunPy \citep{sunpy}. 
Based on the intensity of the X-ray emission in \SI{}{W/m^2}, it is possible to define a logarithmic scale with which to classify solar flares \citep{noaa_goes_solar_flare}. This scale is composed of five main classes: A, B, C, M, and X with different subclasses based on the strength of the flare. The intensity of the X-ray emission of an A-class flare is less than $10^{-7} \ \SI{}{W/m^{2}}$, that of a B-class flare is $10^{-7} - 10^{-6} \ \SI{}{W/m^{2}}$, and that of an X-class flare is of $10^{-4}$ \ \SI{}{W/m^{2}} or more.

\subsection{Heliophysics Events Knowledgebase (HEK)}

The HEK \citep{HEK} is a platform developed to better organise and make more efficient use of the data in the heliophysics field.
We used the HEK to obtain all the peak times of flaring events from 2011 to 2019.

\subsection{Data selection}

As described in Section \ref{sec:introduction}, the purpose of this study is to investigate whether we can train a model to generate high-energy flares filling the lower populated classes with synthetic solar images and thus creating a balanced data set.
Therefore, we do not intend to build a generative model capable of arbitrarily generating an image of the sun characterised by random activity; rather, we aim to generate images of the Sun with a particular class of flare. In order to accomplish this, we must provide our model with data that depict flaring events and are labelled so that the intensity of the flaring event can be determined.

To generate this data set, we proceeded in three steps. First, we setup the access to the HEK flaring events tabular dataset. Although we use all SDOMLv2 data from 2011 to 2019, we need to take into consideration the time gap of 6 minutes between each image (as opposed to 12 seconds in the original SDO data).  Using the SunPy \citep{sunpy} library to connect to the HEK, we first retrieved all flaring events from 2011 to 2019, which total 107,709 distinct flaring events. 

The second step was to associate the HEK flare events with their GOES class. Of the 107,709 HEK flaring events, 15,696 are already associated with a flare class. For the remaining events, we first used SunPy to access the X-ray emission values recorded by GOES/XRSB along with the time of observation. We then associated each flare peak time of the HEK event with the GOES X-ray emission  closest in time, so that the values from GOES are always recorded within a time range of less than 6 seconds after the flare. We used the time after the peak because the decay of the X-ray emission after the flare is less steep than the increase prior to the flare, resulting in more accurate information. 
Up to this point, we had a dataset of 51,374 events characterised by flare class based on the X-ray emission value, the peak time of the flare, and the observation time of the GOES emission.

The third step was to correlate the HEK/GOES associated information with the SDOMLv2 data. We can obtain the observation time of each AIA image, allowing us to link the two datasets. Indeed, for every flaring peak time, we associated the closest image in time such that the image always follows the flare within a 7 minute tolerance. The tolerance of 7 minutes is based on the fact that the time delay from SDOMLv2 is 6 minutes, and we want to maximise the number of images while ensuring the most accurate labelling possible due to data constraints.

As a result, we finally obtained a new set of 20,420 AIA images that are precisely labelled with their GOES flare class. We do not include images without flaring events, because we want to simulate and let the model understand the configuration of these high-energy events at all the levels. In addition, we use full-disc images because we want to test whether or not the model is able to find location of the activity thanks to the attention and convolution layers present in the diffusion model backbone as described in Section \ref{sec:experiments}.

\label{sec:data_sel}
\subsection{Limitation of the new dataset}

The most significant limitation of our new dataset is the time delay. This delay is caused by the 6 minute time cadence in SDOMLv2.
All our results account for this, and a future enhancement could be implemented to mitigate this effect. More details and figures are analysed in Appendix \ref{data_limitation}.
\label{sec:dataset}
\section{Background}
\label{sec:back_ddpm}
In this section, we briefly introduce the DDPMs presented in \citep{ho2020} and their extensions to conditional generation with the classifier-free guidance \citep[CFG]{ho2022classifierfree} on which our study is based.

\subsection{Denoising diffusion probabilistic models}

A diffusion model \citep{ho2020, sohldick} is a type of generative model defined by a forward process -- also called diffusion process -- and a reverse process.

The forward process, described in Eq.\ref{forward_proc}, gradually pushes the samples off the data manifold, turning them into noise. This process is a fixed Markov chain, which gradually adds Gaussian noise and is parameterised by a variance schedule $\beta_{1},..., \beta_{T}$ with $\beta_{t} \in (0, 1) \ \forall t$ and $\beta_{1} < \beta_{2} < ... < \beta_{T}$, where $T$ is the total number of steps of the Markov chain:
\begin{equation}
\begin{aligned}
q(x_{1:T}|x_{0}) := \prod_{t=1}^{T}q(x_{t}|x_{t-1}), \\
q(x_{t}|x_{t-1}) := \mathcal{N}(x_{t}; \sqrt{1-\beta_{t}}x_{t-1}, \beta_{t}\textbf{I}),
\end{aligned}
\label{forward_proc}
\end{equation}
where $\mathcal{N}(a; c)$ denotes a Gaussian distribution with mean 'a' and covariance matrix 'c'.
In the limit of $T$ approaching infinity $q(x_{T}|x_{0}) \sim \mathcal{N}(0, I)$, where $I$ is the identity matrix.
The objective of the model is to determine
\begin{equation}
    p_{\theta}(x_{0}) := \int_{}p_{\theta}(x_{0:T})dx_{1:T},
    \label{model_p}
\end{equation}
$p_{\theta}(x_{0:T})$ is the reverse process defined as:
\begin{equation}
\begin{aligned}
p_{\theta}(x_{0:T}) := p(x_T)\prod_{t=1}^{T}p(x_{t-1}|x_{t}), \\
p(x_{t-1}|x_{t}) := \mathcal{N}(x_{t}; \mu_{\theta}(x_{t}, t), \Sigma_{\theta}(x_{t}, t)),
\end{aligned}
\label{reverse_proc}
\end{equation}
where $p(x_T) = \mathcal{N}(x_{T}, 0, \textbf{I})$, $\mu_{\theta}(x_{t}, t)$ is the predicted mean and $\Sigma_{\theta}(x_{t}, t)$ is the predicted covariance matrix. The reverse process is trained to produce the trajectory back from noise to the data manifold. 

In order to calculate Eq. \ref{model_p}, we have to marginalise over all the possible trajectories $dx_{1:T}$, which is intractable in this form; however we can optimise a variational lower bound on the negative log-likelihood:
\begin{align*}
    \mathbb{E}[-log(p_{\theta}(x_{0}))] \leq \mathbb{E}\left[-log \frac{p_{\theta}(x_{0:T})}{q(x_{1:T}|x_{0})}\right] = \\
    = \mathbb{E}_{q} \left[-log(p(x_{T})) - \sum_{t\geq 1}log\frac{p_{\theta}(x_{t-1}|x_{t})}{q(x_{t}|x_{t-1})}\right] := L,
\end{align*}
which can be rewritten as:
\begin{multline*}
\mathbb{E}_{q}\left[D_{KL}(q(x_{T}|x_{0})||p(x_{T})) + \right. \\ 
\left. + \sum_{t>1}D_{KL}(q(x_{t-1}|x_{t}, x_{0} || p_{\theta}(x_{t-1}|x_{t})) - \log(p_{\theta}(x_{0}|x_{1}))\right],
\end{multline*}
where $D_{KL}(\cdot || \cdot)$ denotes the Kullback–Leibler (KL) divergence.
Working with Gaussians, the KL divergences in the previous equation can be calculated in closed form; as suggested in \customcite{ho2020}, we use the reverse process pasteurisation:
\begin{equation}
    \mu_{\theta}(x_{t}, t) = \frac{1}{\sqrt{\alpha_{t}}}\left(x_{t} - \frac{\beta_{t}}{\sqrt{1-\bar{\alpha}_{t}}}\epsilon_{\theta}(x_{t}, t)\right),
\end{equation}
where $\alpha_{t} = 1 - \beta_{t}$ and $\bar{\alpha}_{t} = \prod_{s=1}^{t}\alpha_{s}$. \\
In conclusion, as in \customcite{ho2020}, we treat the covariance matrix of Eq. \ref{reverse_proc} as a fixed hyper-parameter and we work on the mean, resulting in a simplified loss function:
\begin{equation}
    L(\theta) = \mathbb{E}_{x_{0}, \epsilon \sim \mathcal{N}(0, I)}\left[ || \epsilon_{\theta}(\sqrt{\bar{\alpha}_{t}}x_{0} + \sqrt{1-\bar{\alpha}_{t}}\epsilon, t) - \epsilon ||^{2}_{2} \right]
\end{equation}
where $t \sim \mathcal{U}({1, ...,T})$, $\mathcal{U}$ is the uniform distribution, $\epsilon$ is the noise added to the image in the forward process and $\epsilon_{\theta}$ is the noise predicted by the model.

\subsection{Classifier-free guidance}

Classifier-free guidance was introduced by \customcite{ho2022classifierfree} to ease the process of conditioning the generation models. It has the same effect as classifier guidance \citep{classifierguidance}, but requires no training of a classifier. \\
The target is to change the $\epsilon_{\theta}$ into
\begin{equation}
    \hat{\epsilon}_{\theta}(x_{t}, c) = (1 + w)\epsilon_{\theta}(x_{t}, c) - w\epsilon_{\theta}(x_{t}),
\end{equation}
where $w$ is the CFG scale and $\hat{\epsilon}_{\theta}$ is the conditioned noise predicted as a linear interpolation between the guided prediction and the unguided prediction; the guidance is denoted by $'c'$.
As a result, the model is jointly trained with and without conditions based on a probability set as a hyperparameter, as described in Algorithm 2 of \customcite{ho2022classifierfree}.
\label{sec:background}
\section{Methodology and experiments}

\label{sec:experiments}

The main aim of these experiments is to find the most adapted setup and labelling system to generate full-disc solar images that can be used for further scientific studies and downstream applications. 
The generated synthetic solar images should feature a flare that corresponds to the class specified by a label.

The backbone of the architecture is a DDPM \citep{ho2020}. 
The DDPM consists of a U-Net \citep{ronneberger2015unet}, which is an encoder-decoder network with skip connections where the input and the output shapes are the same. More details on the architecture are given in Appendix \ref{model_arch}. We train for a total of 500 epochs using the AdamW \citep{loshchilov2019decoupled} optimiser, the mean square error (MSE) loss function, a learning rate of $3*10^{-4}$, a batch size of 12 and one NVIDIA TITAN X graphics processing unit (GPU).

To better visualise the performance of the training process, we use the peak signal-to-noise ratio (PSNR), as an evaluation metric.

The model is implemented with the PyTorch framework \citep{pytorch}. The image resolution is 64x64 pixel for computational constraints, although we trained a DDPM with an image size of 128x128 pixel to examine the capabilities of the model if we increase the detail. More information on this experiment is given in Appendix \ref{128x128}.

We trained three models, all of them conditioned to control the amount of generated solar activity present in the image. With this strategy, the specific flare information is encoded in the model, because we train it with this specific supervision.
The distinction between the three models is based on the way we condition (or guide) them:
\begin{itemize}
    \item Discrete labels: GOES classes A, B, C, M, and X,
    \item Continuous labels: GOES X-ray emission value,
    \item Latent space features of an encoder.
\end{itemize}
\label{sec:metrics_analysis}


For the first model, we train the DDPM with the CFG technique as explained in Section \ref{sec:background}. This is straightforward because every image of the dataset is labelled by one of the GOES classes: A, B, C, M and X (Section \ref{sec:data_sel}).

\begin{figure}
\centering
 \includegraphics[width=\hsize]{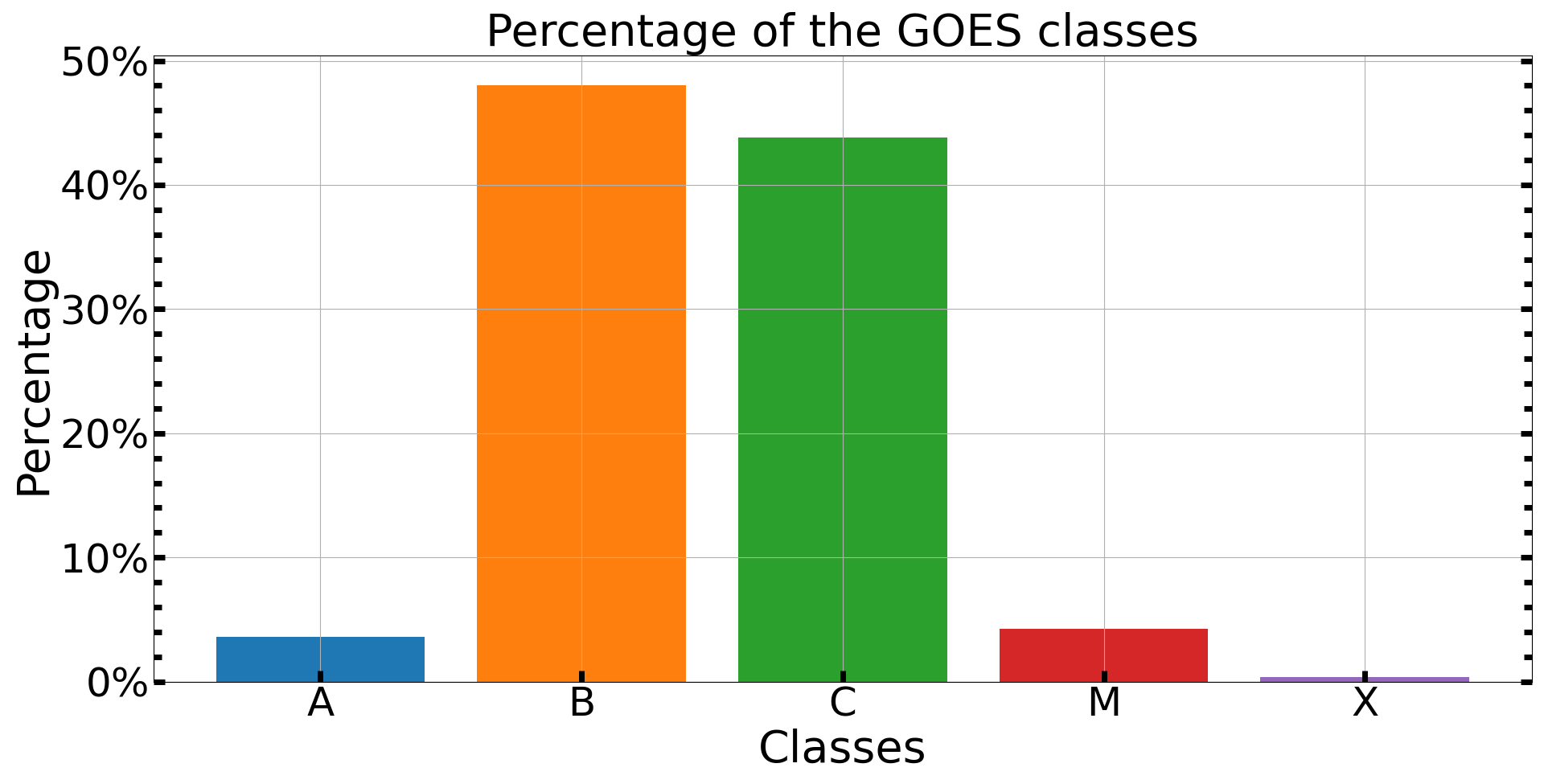}%
\caption{Histogram distribution of the labelled dataset with the discrete GOES labels: A, B, C, M and X.}
\label{fig:perc_goes_clas}
\end{figure}

The data distribution in Figure \ref{fig:perc_goes_clas} is different from the natural distribution of the occurrence of solar flares, where flare distribution functions are successfully modelled using tapered power-law or gamma-function distributions \citep{sakurai2022}. This is not due to how we select data in SDOMLv2, but rather, on the one hand, to the instrumental effect of the GOES spacecraft and, on the other hand, to the threshold of the HEK catalogue, because A-class flare emission is similar to the background emission and so they are not registered as flaring events. To guide the generation and encode the label information, we use an embedding layer, where the size of the dictionary of embeddings is equal to the number of discrete classes and the size of each embedding vector is equal to the size of the time-step embeddings.

\label{discriteguidance}


For the second model, we guide the diffusion directly with the X-ray continuous values obtained from the GOES spacecraft, as explained in Section \ref{sec:data_sel}. This strategy is designed to teach the model the differences between flares of different classes, avoiding the somewhat arbitrary repartition of flares into classes (e.g. a large B flare is more similar to a small C flare than to a small B flare).
This way, we are able to better parameterise the class boundaries. 

To guide the generation, we take the X-ray value, encode it with a sequence of two linear layers up to when the dimensions of the value are the same as those of the time-step embedding and then we sum them up. 


For the third model, we guide the diffusion with the discrete labels as 
in our first model, but wee also add the feature embeddings of a context-encoder variational autoencoder (ceVAE) already pretrained on SDO data \citep{marius, zimmerer2018contextencoding}. The ceVAE architecture combines a Context Encoder (CE) and a Variational Autoencoder (VAE). A CE is a type of deep learning model that is trained to reconstruct an input image after randomly masking local patches of it. On the other hand, a VAE is a type of generative model that simultaneously learns the representation of the input data and its probabilistic distribution. The VAE assumes that the distribution of the latent space is Gaussian. As analysed in \customcite{zimmerer2018contextencoding}, the CE and the VAE are trained together, and share the same weights for the encoder-decoder architecture. 

\begin{figure}
    \centering
    \includegraphics[width=\hsize]{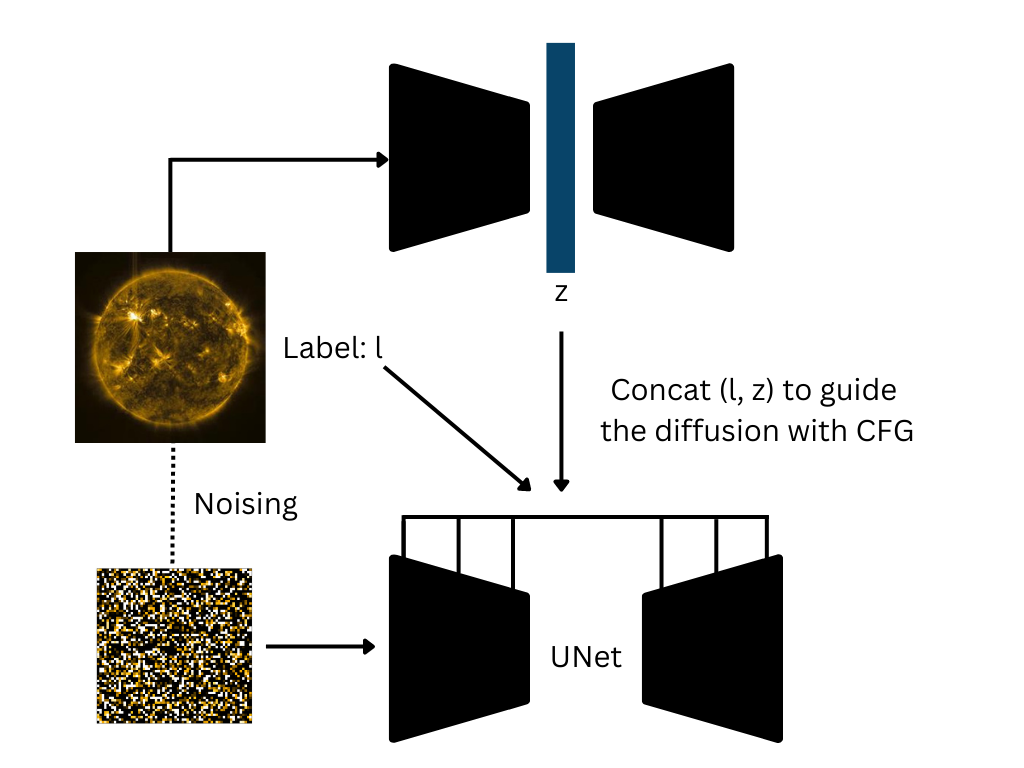}
    \caption{Sketch of the network trained with the discrete labels and the ceVAE embeddings to guide the diffusion. In the concatenation process l represents the discrete label and z the features of the ceVAE latent space.}
    \label{fig:ceVAEembnet}
\end{figure}
A sketch of our network is presented in Figure \ref{fig:ceVAEembnet}.
This procedure is designed to prepare the DDPM, giving it compressed information on the Sun with a specific amount of activity.

To guide the generation, we rely on a ceVAE that has been pretrained on SDO data, and train a DDPM to generate new ceVAE-like embeddings containing compressed information on the Sun with a specific level of activity, as described in Dall-E 2 \citep{ramesh2022}.

In conclusion, we use the ceVAE architecture as a baseline for our three models. The training details and performance of this architecture can be found in \customcite{marius}.

\section{Metrics}

\label{metr_4_1}
The metrics involved in determining the quality of our models are the cluster metrics \citep{hackstein2023}, which evaluate whether or not the generative model can produce data with the same distribution as true data without mode collapse, the Fréchet inception distance (FID) \citep{FID}, which determines the quality of generated images as well as the quality of the generated distribution; and the F1 score \citep{f1score}.

The F1 score is calculated based on the precision and recall of a classification model. The precision is the number of true-positive predictions divided by the total number of positive predictions, while the recall is the number of true positive predictions divided by the total number of true-positives in the dataset. In the context of image comparison, precision and recall can be thought of as measures of how accurately the generated image matches the true image. The F1 score combines these measures into a single value, which provides an overall assessment of the similarity between the two images. Therefore, the use of the F1 score in this context is designed to quantitatively evaluate how well a generated image matches a true image with the same amount of activity (e.g. to determine whether or not a generated image with an X flare is similar to a true image with an X flare), based on the precision and recall of the classification model used to make the comparison. Thus, to evaluate how similar a generated image is to a ground-truth image of the same class, we train a supervised classifier on true data, test it on generated data looking at the F1 score per class, and then we take the macro F1 score. 
\section{Results}
\label{sec:results}
\begin{table*}
    \centering
    \begin{tabularx}{\textwidth}{|l|X|X|X|X|}
    \hline
        \textbf{Metric} & \textbf{ceVAE (baseline)} & \textbf{Discrete (ours)} & \textbf{Continous (ours)} & \textbf{ceVAE\_Emb (ours)} \\ \hline
        Cluster error GT & \multicolumn{4}{c|}{0.00197} \\ \hline
        Cluster distance GT & \multicolumn{4}{c|}{1.00104} \\ \hline
        Cluster Std GT & \multicolumn{4}{c|}{0.99816} \\ \hline
        Cluster error GEN  $\downarrow$ & 7.9478 ± 0.9137 & \textbf{0.1294 ± 0.0358} & 1.5031 ± 0.1476 & 0.2073 ± 0.0361 \\ \hline
        Cluster distance GEN $\downarrow$ & 2.2057 ± 0.0096 & 0.9212 ± 0.0037 & \textbf{0.9342 ± 0.0023} & 0.8377 ± 0.0055 \\ \hline
        Cluster Std GEN $\downarrow$ & 3.2382 ± 0.0096 & 1.2107 ± 0.0037 & \textbf{1.0976 ± 0.0023} & 1.4801 ± 0.0055 \\ \hline
        FID CLIP $\downarrow$ & 5.05 & 0.122 & \textbf{0.057} & 0.39 \\ \hline
        FID IV3 $\downarrow$ & 215.933 & 3.693 & \textbf{2.703} & 12.264 \\ \hline
        F1 score $\uparrow$ & ~ & \textbf{0.7} & 0.34 & 0.6 \\ \hline
        Precision $\uparrow$ & ~ & \textbf{0.73} & 0.35 & 0.6 \\ \hline
        Recall $\uparrow$ & ~ & \textbf{0.74} & 0.37 & 0.7 \\ \hline
    \end{tabularx}
    \caption{Results of the experiments based on the metrics of Section \ref{sec:experiments}. The symbol $\downarrow$ indicates that a lower value is preferable for the metric it represents, while the symbol $\uparrow$ indicates that a higher value is preferable for that metric. The F1 score, precision, and recall are designed such that their maximum value is 1.}
    \label{results_metrics}
\end{table*}

The results of these experiments are shown in Table \ref{results_metrics}.
We produced a total of 60,000 images for these analyses, with each of the three models contributing 20,000 samples. Half of these images (30,000) were generated to reflect the proportions of image classes (A, B, C, M and X) found in the original dataset, while the other half (30,000) were generated uniformly with each image class equally represented. Specifically, for each of the three models, the generated images were separated into 20 sets of 1000 images each. In ten of these sets, each category (A, B, C, M, and X) is represented by 200 images (uniformly generated). In contrast, the class distribution in the remaining ten sets mirrors the imbalances evident in the original dataset, where each class is represented according to the percentages shown in Figure \ref{fig:perc_goes_clas}. The reason for generating data with different distributions, as explained before, relies on the metrics used to evaluate the performance of our models. The cluster metrics and the FID indeed measure the similarity between the generated distribution and the true distribution, and so both of them need to be compared on a generated dataset with the same characteristics as the true dataset; in this case, in terms of class percentages. On the other hand, the F1 score, the precision and the recall are used to determine if a generated image of a particular class is similar to a true image of the same class, and therefore the trained classifier (see Section \ref{metr_4_1}) should be tested on a uniform dataset without imbalance in the class percentages; otherwise our results would be biased.

\begin{figure*}
\centering
\subfloat[\label{fig:clip}]{%
 \includegraphics[width=0.45\linewidth]{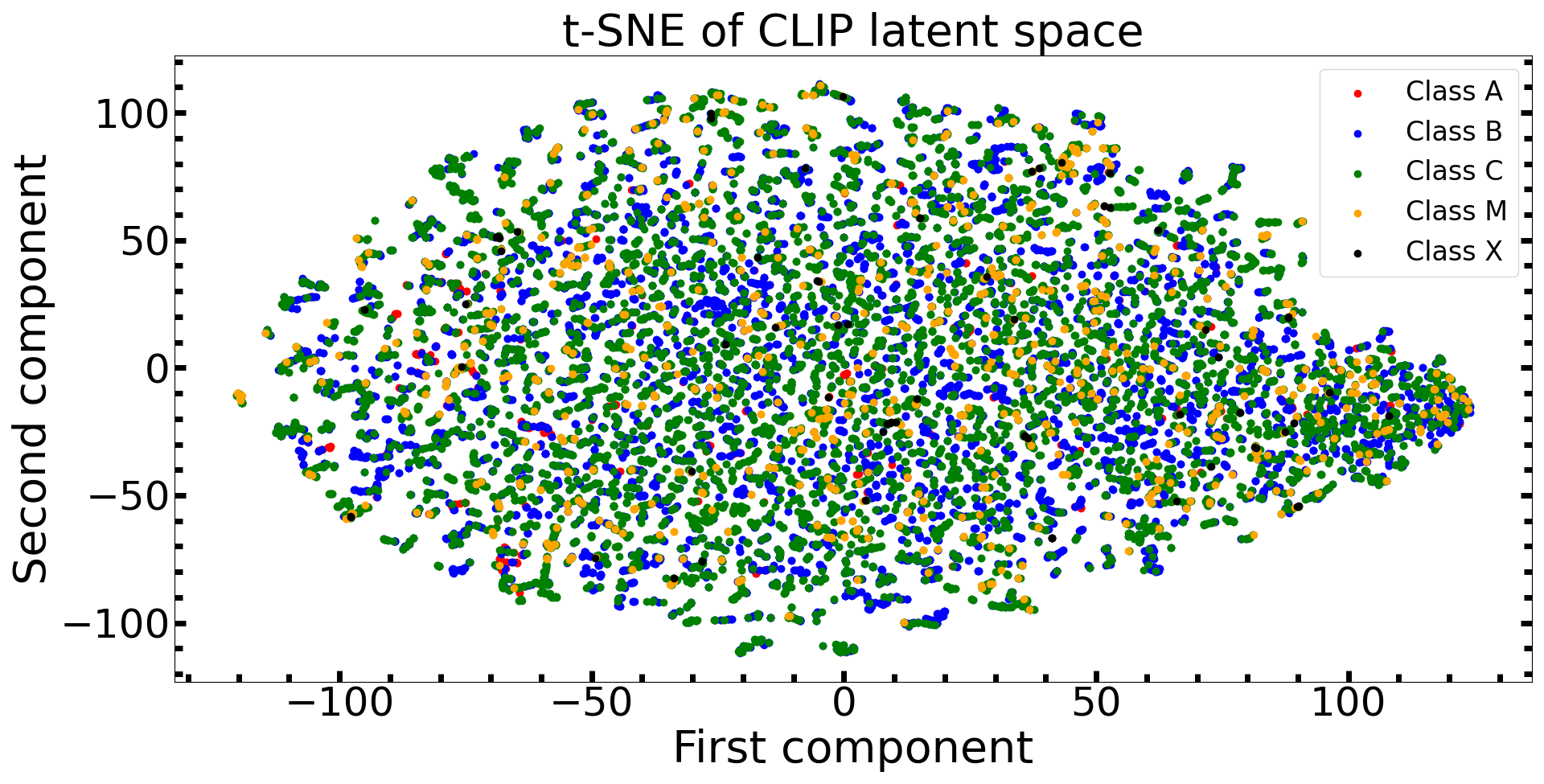}%
}\hspace{0.05\linewidth}
\subfloat[\label{fig:ceVae}]{%
 \includegraphics[width=0.45\linewidth]{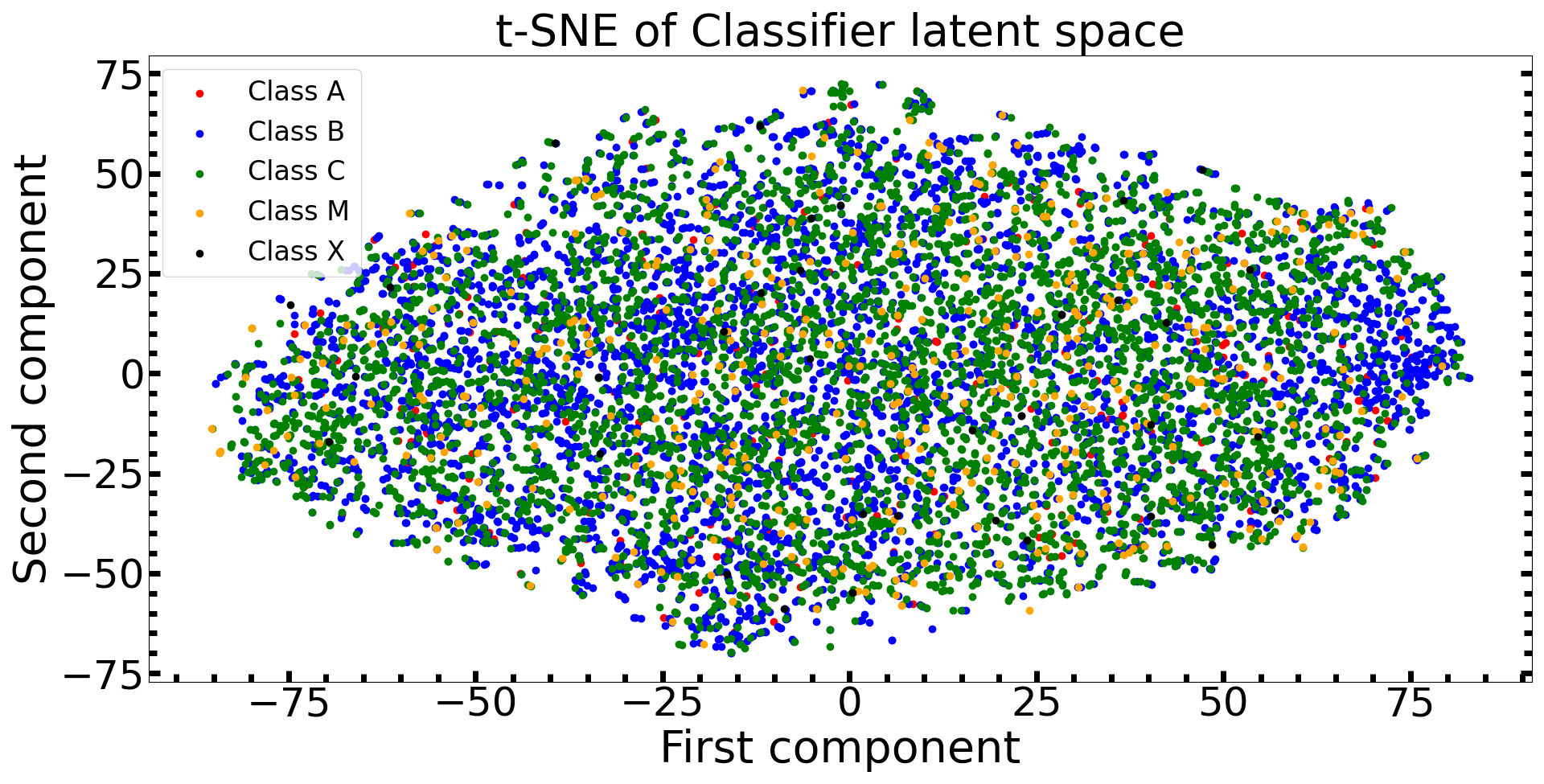}%
}
\par\bigskip
\subfloat[\label{fig:classifier}]{%
 \includegraphics[width=0.6\linewidth]{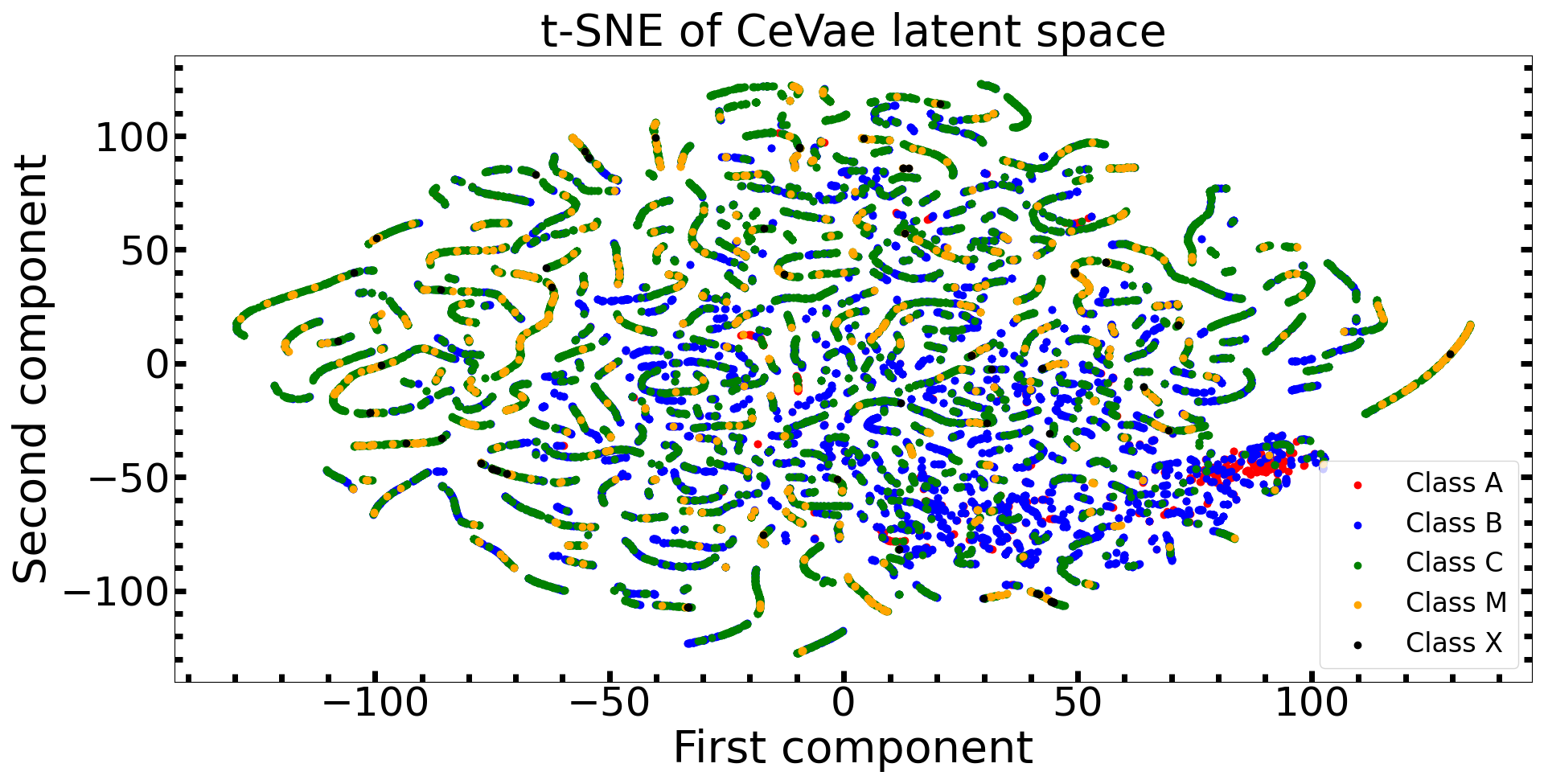}%
}
\caption{t-SNE dimensionality-reduction technique applied to various latent spaces to determine which is most appropriate for cluster metrics. Figure a) shows the t-SNE of CLIP latent space. Figure b) shows the t-SNE of the latent space of a classifier. Figure c) shows the latent space of a pretrained ceVAE.}
\label{fig:latentspace}
\end{figure*}

Furthermore, for the cluster metrics we need a latent space to compute the calculations. For this reason, we decided to analyse different feature spaces using the t-SNE dimensionality-reduction technique \citep{vanDerMaaten2008}.
The 512-dimensional ceVAE \citep{marius} latent space is the most accurate representation of the class division (A, B, C, M, X); as shown in Figure \ref{fig:latentspace}, it is the only latent space where clustering can be inferred, and the filamentous structure is related to the images that are close in time and thus very similar, underlying a major completeness with respect to the CLIP latent space and the classifier latent space. The classifier has been trained in a supervised manner, and the lack of distinct clusters in the latent space can be attributed to insufficient data in different classes. In such cases, it has been demonstrated that unsupervised methods can outperform supervised models  \citep{slavaolga}.

\begin{figure*}
    \begin{minipage}{0.5\textwidth}
        \centering
        \begin{subfigure}{\linewidth}
            \includegraphics[width=\linewidth]{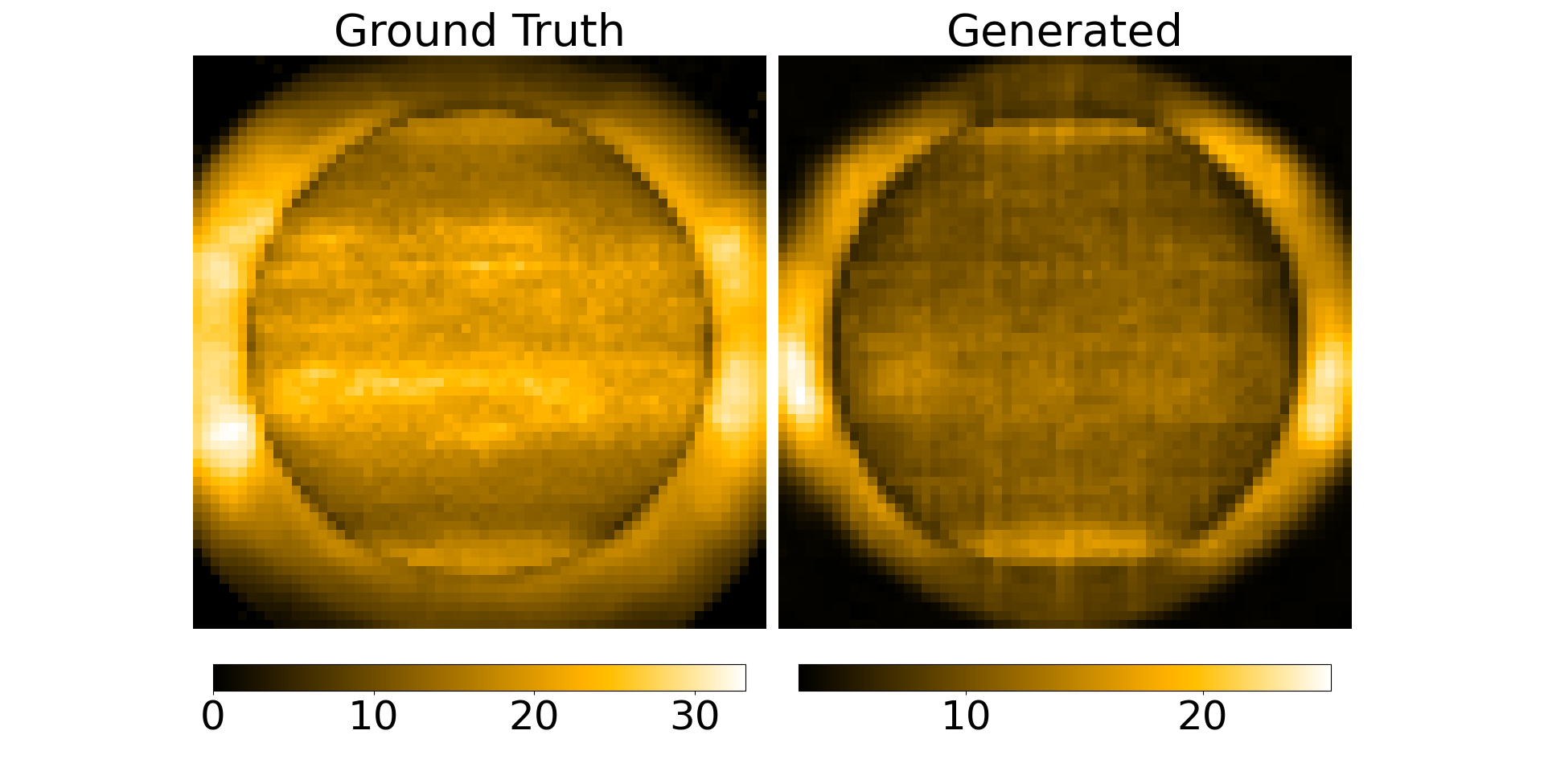}
            \caption{}
            \label{fig:clipA}
        \end{subfigure}
    \end{minipage}
    \begin{minipage}{0.5\textwidth}
        \centering
        \begin{subfigure}{\linewidth}
            \includegraphics[width=\linewidth]{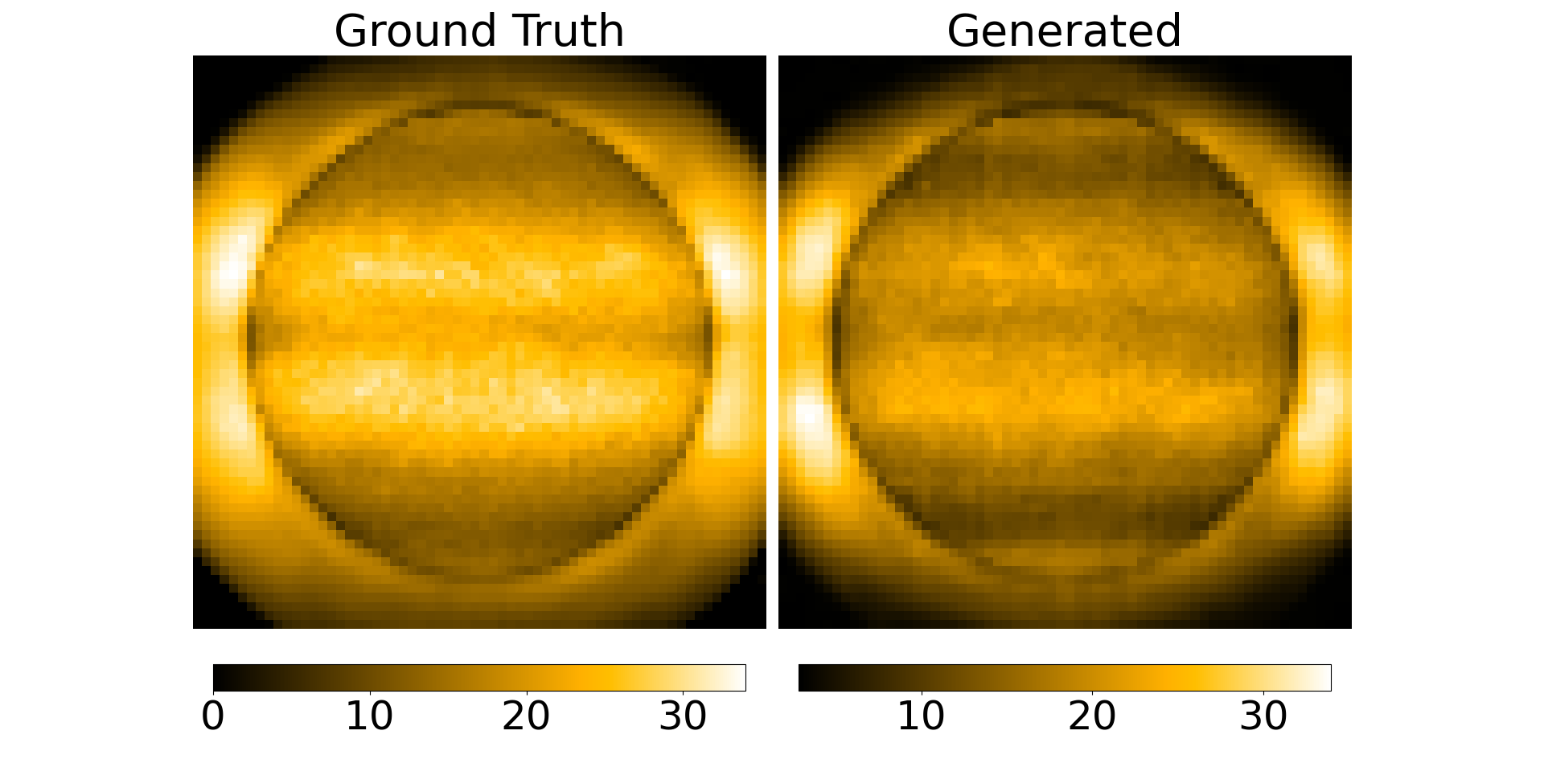}
            \caption{}
            \label{fig:ceVaeB}
        \end{subfigure}
    \end{minipage}
    
    \vspace{\baselineskip} 
    \begin{center}

    \begin{minipage}{0.5\textwidth}
        \centering
        \begin{adjustbox}{valign=M,center} 
            \begin{subfigure}{\linewidth}
                \includegraphics[width=\linewidth]{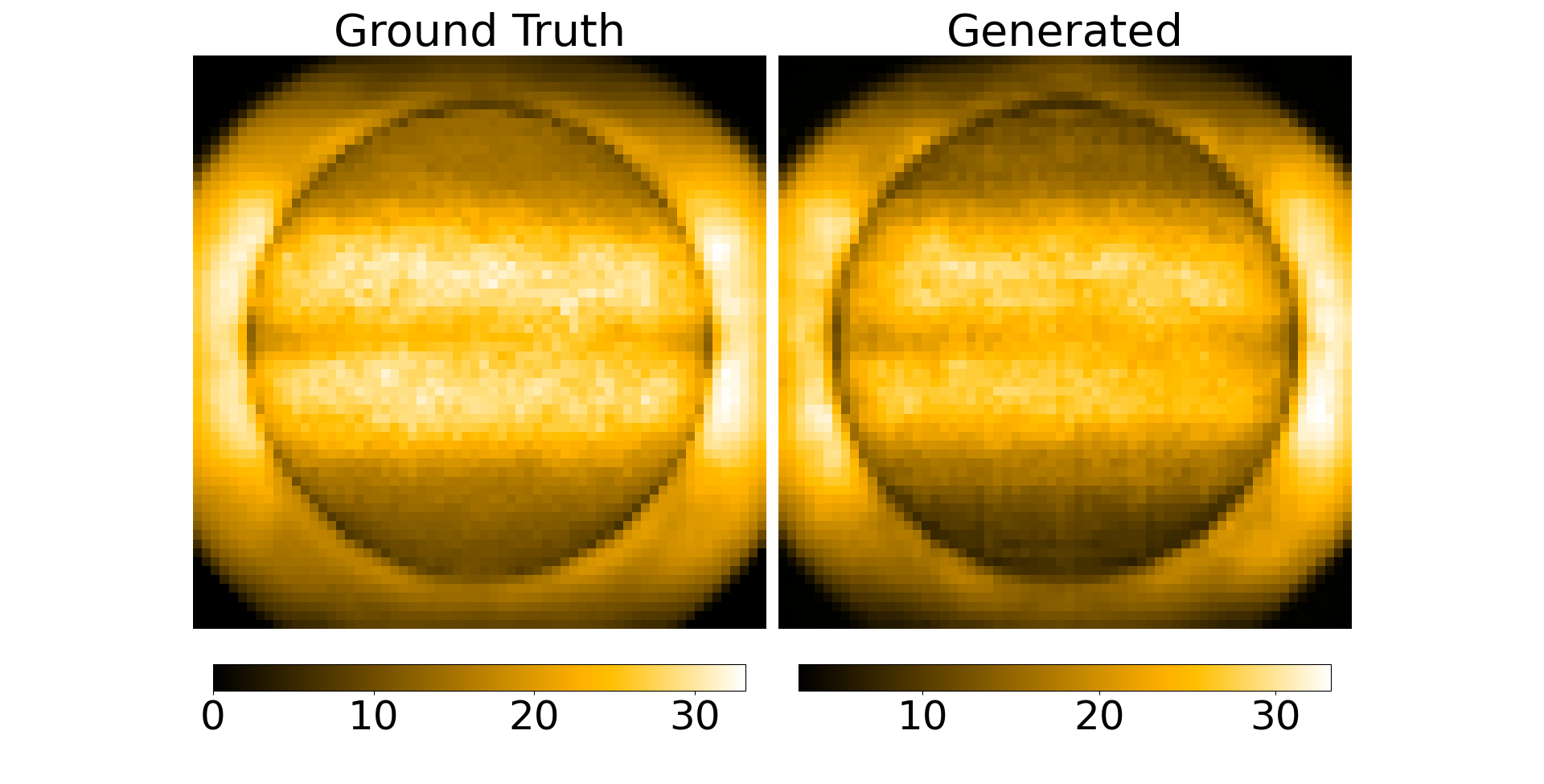}
                \caption{}
                \label{fig:classifierC}
            \end{subfigure}
        \end{adjustbox}
    \end{minipage}
    \end{center}
     
    \vspace{\baselineskip} 
    
    \begin{minipage}{0.5\textwidth}
        \centering
        \begin{subfigure}{\linewidth}
            \includegraphics[width=\linewidth]{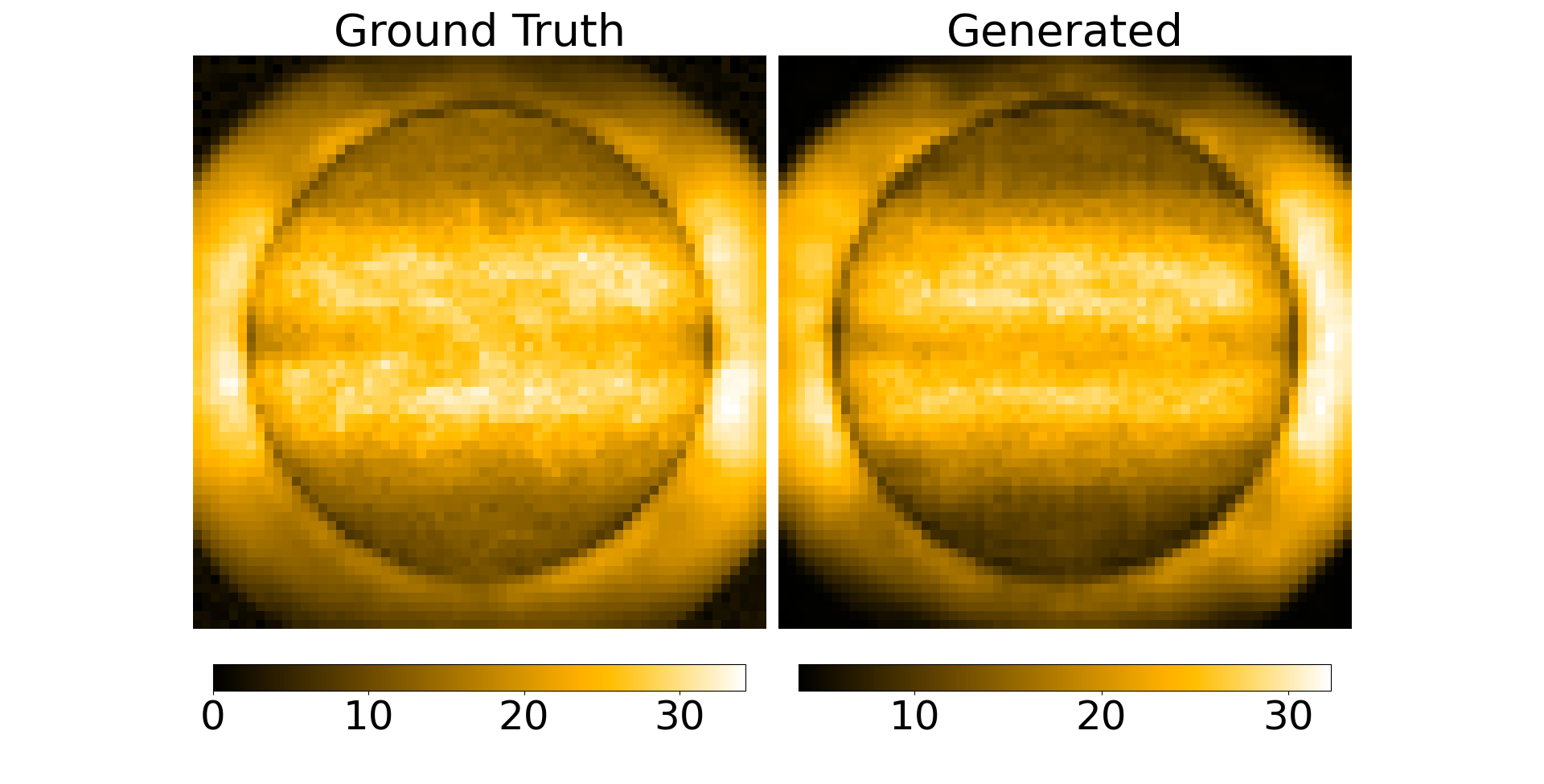}
            \caption{}
            \label{fig:classifierM}
        \end{subfigure}
    \end{minipage}
    \begin{minipage}{0.5\textwidth}
        \centering
        \begin{subfigure}{\linewidth}
            \includegraphics[width=\linewidth]{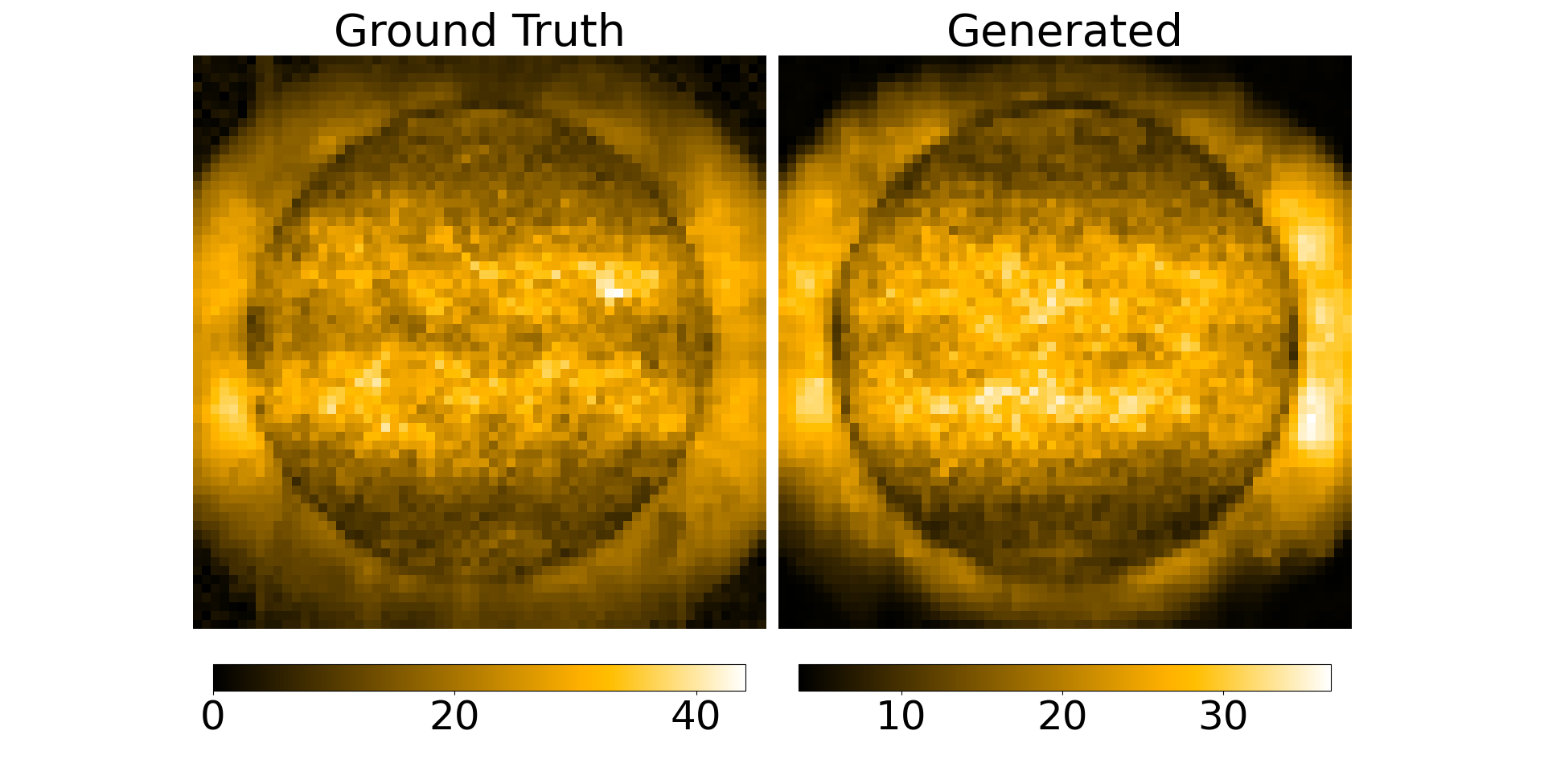}
            \caption{}
            \label{fig:classifierX}
        \end{subfigure}
    \end{minipage}
    
    \caption{Standard-deviation maps, comparing true images (left) and generated images (right) for each class. Panel a) represents the A class, panel b) the B class, panel c) the C class, panel d) the M class and panel e) the X class.}
    \label{fig:stdmap}
\end{figure*}

\begin{figure*}
\centering
\subfloat[\label{fig:simexample}]{%
 \includegraphics[width=0.14\hsize]{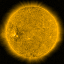}%
}
\qquad
\subfloat[\label{fig:obsexample}]{%
 \includegraphics[width=0.14\hsize]{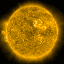}%
}
\qquad
\subfloat[\label{fig:obsexample}]{%
 \includegraphics[width=0.14\hsize]{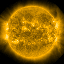}%
}
\qquad
\subfloat[\label{fig:obsexample}]{%
 \includegraphics[width=0.14\hsize]{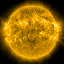}%
}
\qquad
\subfloat[\label{fig:obsexample}]{%
 \includegraphics[width=0.14\hsize]{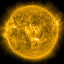}%
}
\qquad
\subfloat[\label{fig:obsexample}]{%
 \includegraphics[width=0.14\hsize]{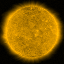}%
}
\qquad
\subfloat[\label{fig:obsexample}]{%
 \includegraphics[width=0.14\hsize]{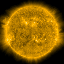}%
}
\qquad
\subfloat[\label{fig:obsexample}]{%
 \includegraphics[width=0.14\hsize]{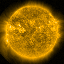}%
}
\qquad
\subfloat[\label{fig:obsexample}]{%
 \includegraphics[width=0.14\hsize]{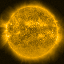}%
}
\qquad
\subfloat[\label{fig:obsexample}]{%
 \includegraphics[width=0.14\hsize]{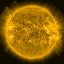}%
}
\qquad
\subfloat[\label{fig:obsexample}]{%
 \includegraphics[width=0.14\hsize]{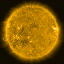}%
}
\qquad
\subfloat[\label{fig:obsexample}]{%
 \includegraphics[width=0.14\hsize]{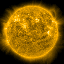}%
}
\qquad
\subfloat[\label{fig:obsexample}]{%
 \includegraphics[width=0.14\hsize]{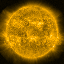}%
}
\qquad
\subfloat[\label{fig:obsexample}]{%
 \includegraphics[width=0.14\hsize]{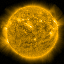}%
}
\qquad
\subfloat[\label{fig:obsexample}]{%
 \includegraphics[width=0.14\hsize]{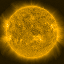}%
}
\qquad
\subfloat[\label{fig:obsexample}]{%
 \includegraphics[width=0.14\hsize]{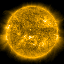}%
}
\qquad
\subfloat[\label{fig:obsexample}]{%
 \includegraphics[width=0.14\hsize]{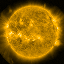}%
}
\qquad
\subfloat[\label{fig:obsexample}]{%
 \includegraphics[width=0.14\hsize]{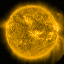}%
}
\qquad
\subfloat[\label{fig:obsexample}]{%
 \includegraphics[width=0.14\hsize]{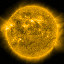}%
}
\qquad
\subfloat[\label{fig:obsexample}]{%
 \includegraphics[width=0.14\hsize]{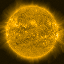}%
}
\qquad
\subfloat[\label{fig:obsexample}]{%
 \includegraphics[width=0.14\hsize]{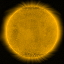}%
}
\qquad
\subfloat[\label{fig:obsexample}]{%
 \includegraphics[width=0.14\hsize]{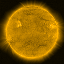}%
}
\qquad
\subfloat[\label{fig:obsexample}]{%
 \includegraphics[width=0.14\hsize]{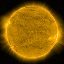}%
}
\qquad
\subfloat[\label{fig:obsexample}]{%
 \includegraphics[width=0.14\hsize]{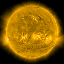}%
}
\qquad
\subfloat[\label{fig:obsexample}]{%
 \includegraphics[width=0.14\hsize]{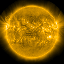}%
}

\caption{Batch of 25 generated images. The first two rows are generated with the discrete label model, the third and the fourth row with the X-ray model and the last row with the ceVAE embedding model. The first column shows the A class, the second column the B class, the third column the C class, the fourth column the M class and the fifth column the X class.}
\label{fig:generated_images}
\end{figure*}

The cluster metrics are based on the K-means unsupervised clustering with the Sklearn library \citep{sklearn_api}.
In Table \ref{results_metrics}, the results of the cluster metrics GEN (generated) should be compared to the values of the cluster metrics GT (ground truth), which serve as a benchmark. As there are five GOES classes A, B, C, M, and X, the number of clusters used for calculating these metrics is five. The model trained with discrete GOES classes has the lowest cluster error, while the model trained with continuous X-ray values has the lowest cluster distance and cluster standard deviation. The cluster error measures whether the clusters in feature space contain the same number of samples as the target distribution. Consequently, this metric has the potential to reveal mode collapse. This means that the model trained with the discrete labels can produce a distribution of data that can be clustered similarly to the true distribution. The cluster distance and standard deviation, on the other hand, determine whether the generated samples populate the correct regions in feature space with sufficient diversity and in this case the best is the model trained with x-ray continuous values.
This suggest that the model trained with discrete labels is better at reproducing the overall structure of the data, while the model trained with continuous X-ray values is better at capturing the finer details and ensuring diversity in the generated samples. Nevertheless, as we can see from Figure \ref{fig:generated_images}, the model trained with the GOES classes better differentiates between the energy classes. Indeed with the X-ray model will not be possible to generate images of the Sun in an extremely calm or extremely active state, resulting in a uniform production of activity across all levels.\\
For the FID, we used the Python library clean-fid \citep{parmar2021cleanfid}. The latent spaces considered for the FID are the ViT-B/32 CLIP encoder and the InceptionV3 encoder \citep{radford2021learning, karras2020training}, specifically for consistency with the literature, so that we can make more meaningful comparisons of our results. More effective models are characterised by lower FID values, and here we therefore find the X-ray model to be the most effective. However, as seen in Figure \ref{fig:generated_images}, the X-ray model is always generating activity and this leads to a lower value of the FID, as before for the cluster metrics. This means that the best is the model trained with the discrete GOES labels even though the FID is slightly higher with respect to the X-ray model. In addition to visual inspection, we can confirm this trend, with the F1 score, the precision and the recall at the end of Table \ref{results_metrics} (these represent the macro values, which are the averages among the classes; the values for each class is given in Appendix \ref{metricsperclass}). As stated in Section \ref{sec:metrics_analysis}, we trained a supervised classifier on the true data using the distilled data-efficient image transformer (DeIT) backbone \citep{touvron2021training} to teach the model how to recognise true A-class, B-class, C-class, M-class image, and X-class images. The value of the F1 score, the precision and the recall on true data are respectively 0.55, 0.57 and 0.54. These benchmark values serve as a reference point. When assessing the performance of our trained classifier on the generated data, we compare the obtained results to those achieved on the true data by dividing the former by the latter. Performing this analysis on the ceVAE model is not feasible because it is not a conditioned model, and therefore, it is not possible to generate an image with a specific flare. Subsequently, we evaluate this classifier on generated data in order to determine which model produces images that are most similar to the actual images for the respective GOES classes. As a result of this analysis, the model trained with discrete GOES classes is the best model in terms of F1 score, precision, and accuracy, with a macro F1 score of 0.38, which is the 70 \% (0.7) of to the best score we can achieve (on true data) 0.54, whereas the X-ray model achieves only the 34 \% (0.34) as macro F1 score of the baseline.
\section{Image Analysis}
\label{sec:analysis}
To the best of our knowledge, our method is the first to generate images of the Sun with the ability to control its activity and the first to apply the novel concept of DDPM \citep{ho2020} to the field of heliophysics. Based on the results of Sect. \ref{sec:results}, the best model in terms of visual inspection, distribution generation (cluster metrics) and applicability (F1 score) is the model trained with the discrete GOES labels: A, B, C, M and X. It is possible to control the presence and intensity of a solar flare on simulated SDO images of the Sun without copying the data from the training set, as evidenced by the fact that the cluster metrics do not perfectly match the reference values, and by the standard deviation maps (std maps) in Figure \ref{fig:stdmap}. For every class the images of the left panel in Figure \ref{fig:stdmap} represent the variation in the true images, whilst those in the right panel the generated images (using the discrete model). To compute the standard deviation maps, we concatenate the images along the batch dimension and then calculate the standard deviation per pixel, so that the brighter regions correspond to regions with greater variation and thus greater activity. We can see that the active regions on generated images are similar in terms of position to the real data but are never in exactly the same part of the image and with the same intensity, even for the X standard deviation maps with only 47 images in the training set. Furthermore, we never observe active regions at the Sun’s poles, which is consistent with physical observations. However, the generated standard deviation maps on A images is the most divergent from the actual data. Indeed, the A-generated images are extremely stable, with few variations, regardless of the fact that the training set contains A images with some activity. Further tests are needed in order to better analyse this phenomenon \citep{somepalli2022diffusion} and we plan to carry on such tests in future works. Given the present findings, we can conclude that the model is able to generate all the types of activity present in the training set, with a minor limitations being its lack of ability to generate A-level images that are nearer to B-level than to low A-level.
\section{Model usage}
\label{sec:discussion}
We now turn to two possible downstream experiments of the best model considered in this study based on the results obtained in Section \ref{sec:results}. This procedure considers the strengths and weaknesses of the model, as well as its ability to generalise. Our ultimate objective is not for the model to beat all existing models on those tasks, but showing that the usage of generated images has a positive impact with respect to not using them.

\subsection{Classification experiment}

Given the results in Table \ref{results_metrics} and  the potential of the DDPM to generate synthetic solar images, we tested whether or not we can use them to overcome the problem of unbalanced data. For this purpose, we trained a supervised classifier with the same architecture as in Section \ref{sec:experiments}; we first did this without the addition of generated data to the least represented classes of the training set, A, M and X, and then we added 200, 400 and 600 synthetic images per represented class, respectively. The aim of this excercise is to see if adding the generated images improves the performance of our classifier, boosting the detection of under-represented classes. In total, we trained four identical supervised classifiers, with the only difference between them being the addition of the generated samples. Naturally, each time, we tested on the same set of real data. 

The proportion of added data is small compared to the size of the entire dataset. Therefore, the dataset remains unbalanced. This is intentional, as the experiment is to measure the impact of adding synthetic images -- even in small numbers. For each incremental addition (200, 400, and 600), we used three distinct sets of generated images to better understand the resulting variations in the obtained values. In other words, for each addition, we trained three separate classifiers with different sets of 200, 400 and 600 images, respectively. This approach allows us to gain insight into the variations that arise from these different additions.
\begin{figure}
    \centering
    \begin{subfigure}{0.3\textwidth}
        \includegraphics[width=\linewidth]{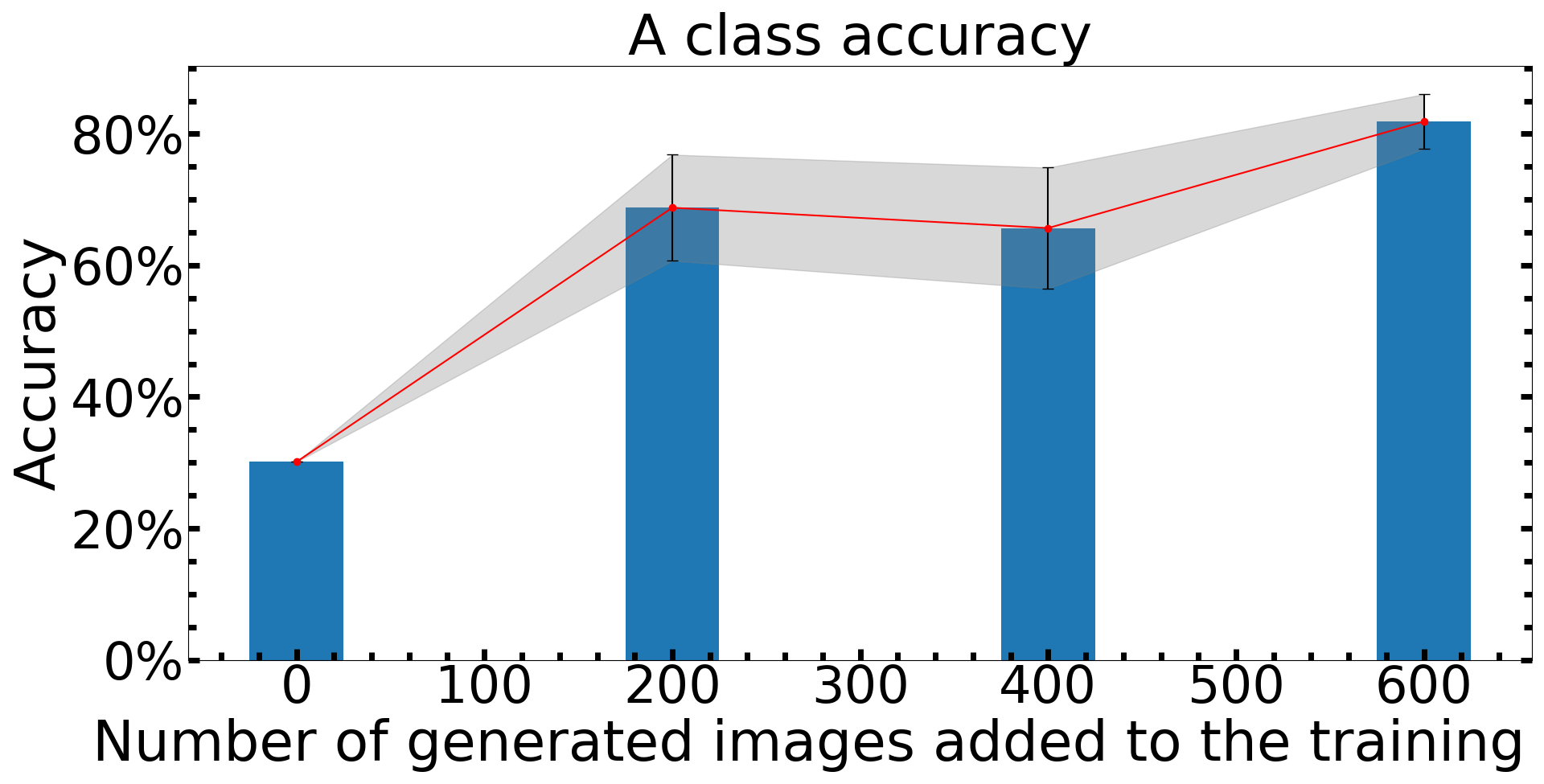}
        \caption{}
        \label{fig:clip}
    \end{subfigure}
    \hfill 
    \begin{subfigure}{0.3\textwidth}
        \includegraphics[width=\linewidth]{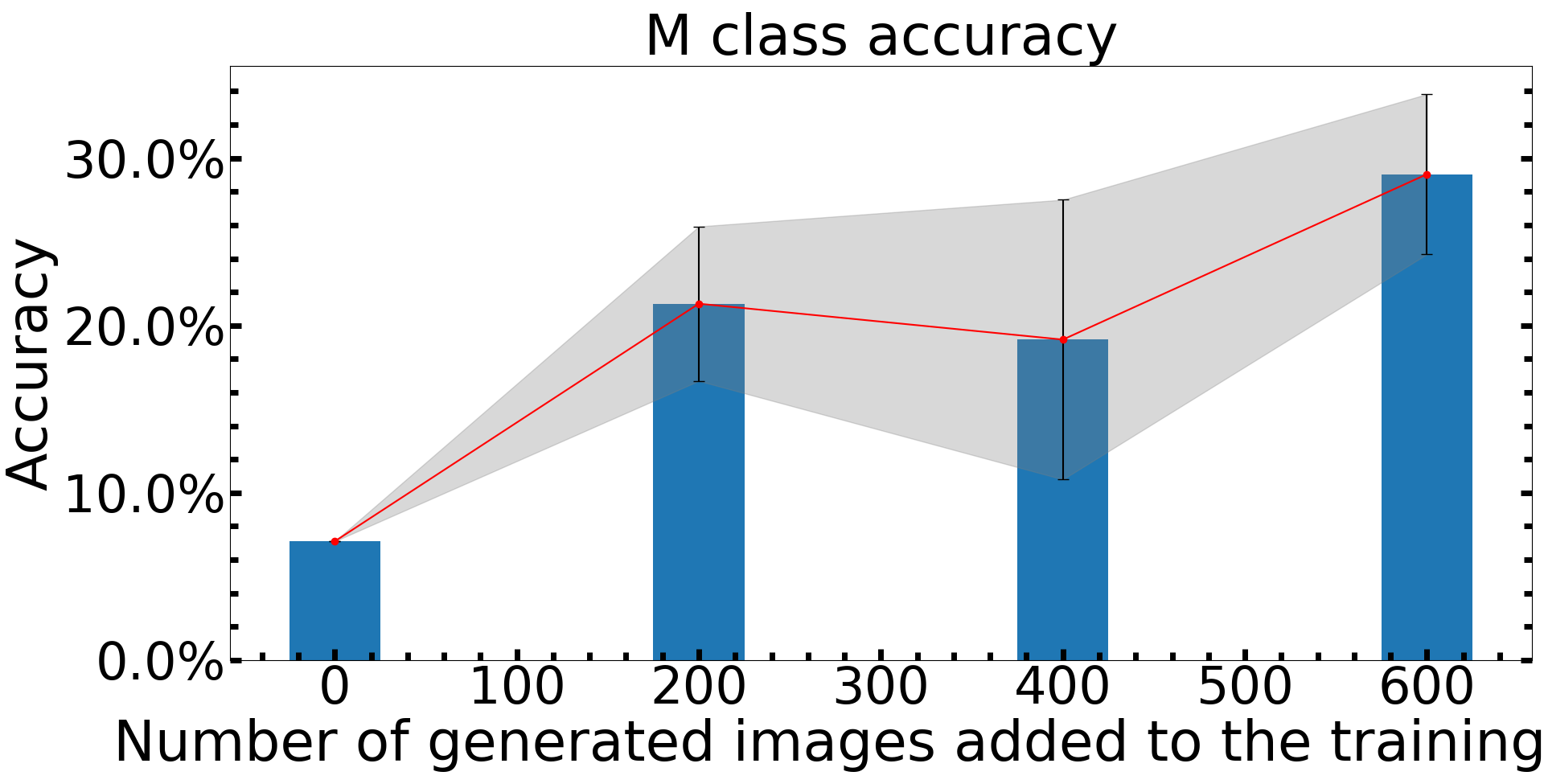}
        \caption{}
        \label{fig:classifier}
    \end{subfigure}
    \hfill 
    \begin{subfigure}{0.3\textwidth}
        \includegraphics[width=\linewidth]{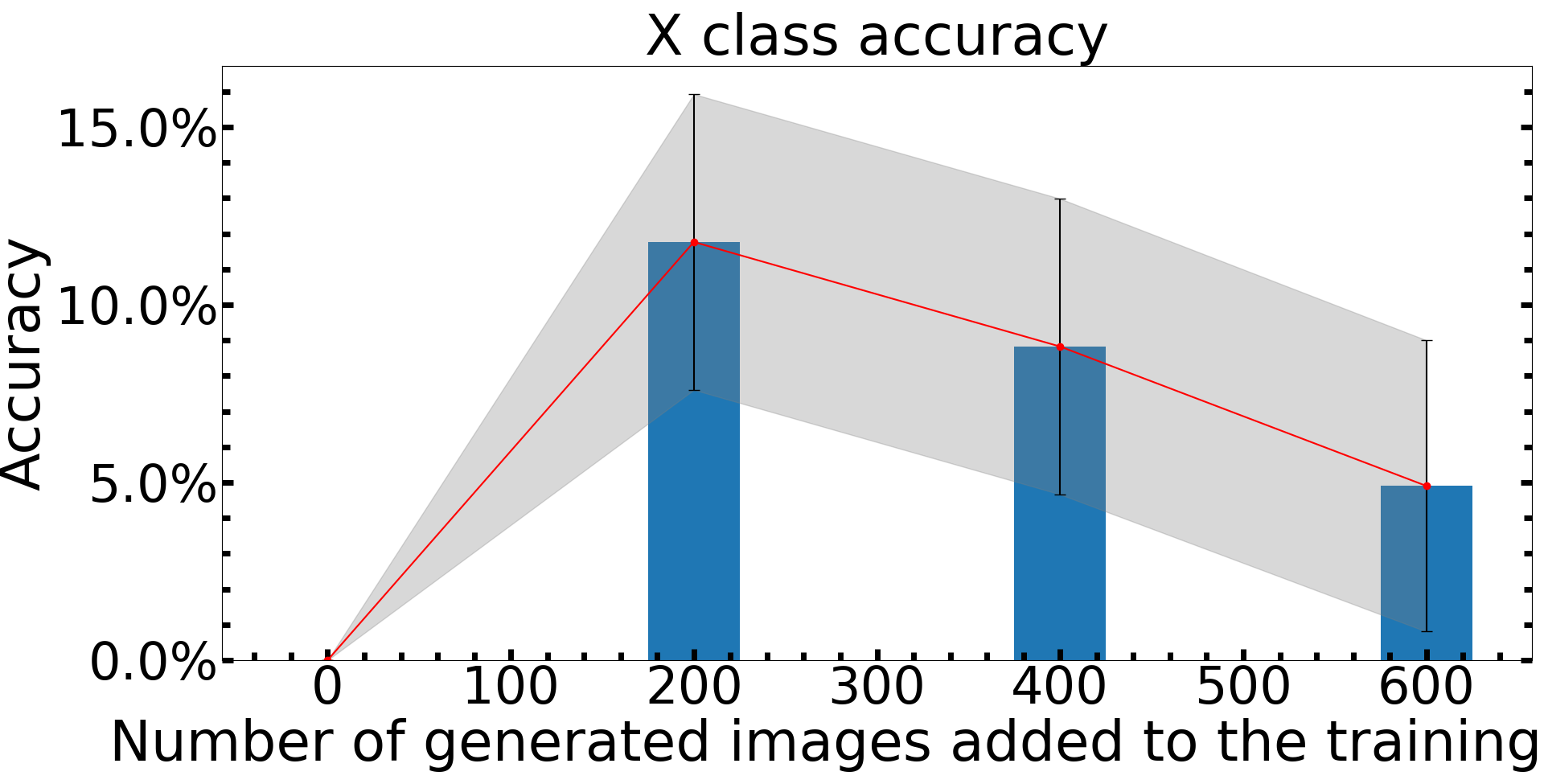}
        \caption{}
        \label{fig:classifier}
    \end{subfigure}
    \caption{Variation of accuracy per class increasing the number of generated samples added. Panel a) shows the evolution of the accuracy of A class, panel b) shows that of M class and panel c) shows that of X class.}
    \label{fig:accuracyvar}
\end{figure}

\begin{figure}
    \centering
    \begin{subfigure}{0.3\textwidth}
        \includegraphics[width=\linewidth]{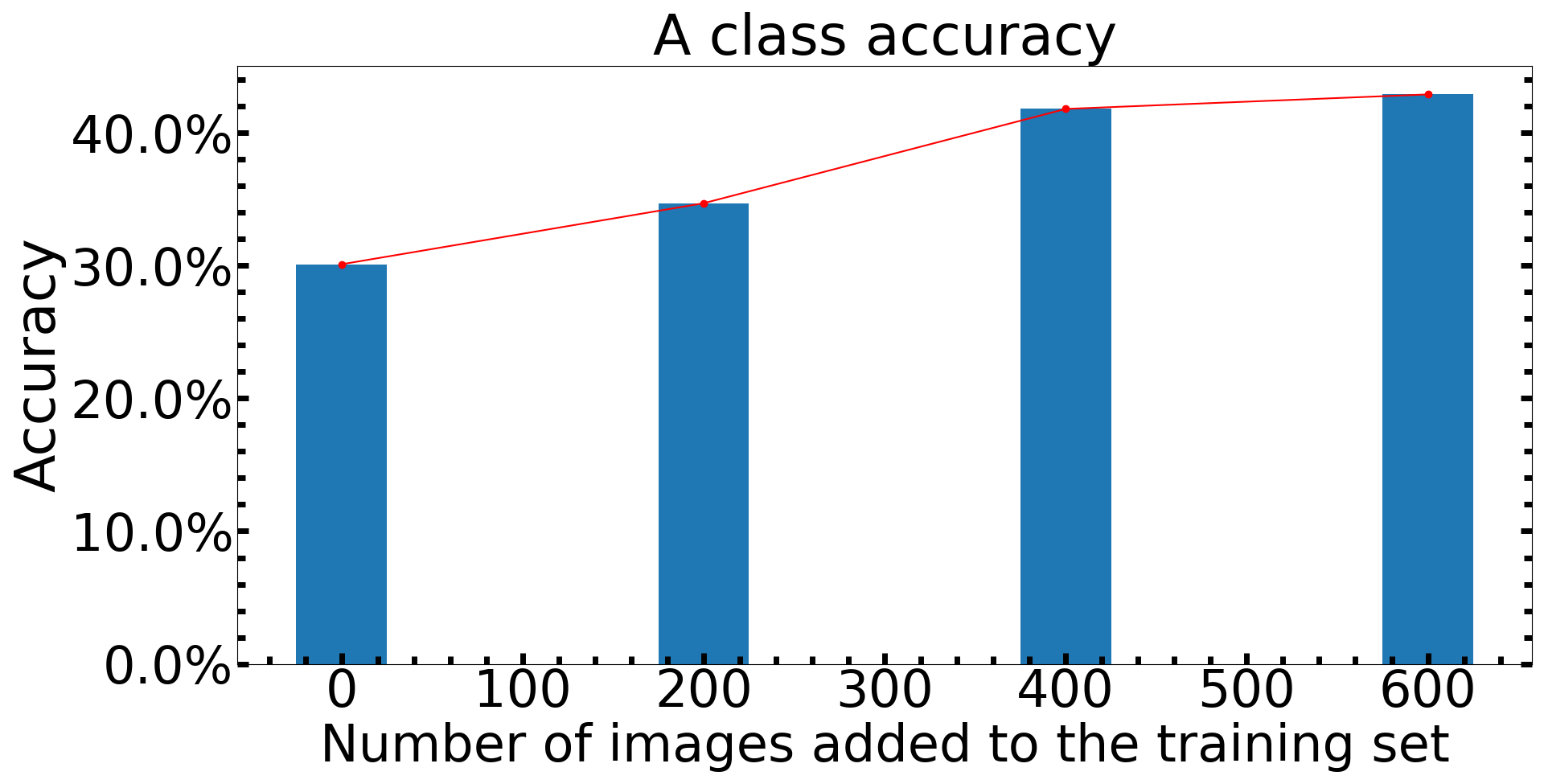}
        \caption{}
        \label{fig:clip}
    \end{subfigure}
    \hfill 
    \begin{subfigure}{0.3\textwidth}
        \includegraphics[width=\linewidth]{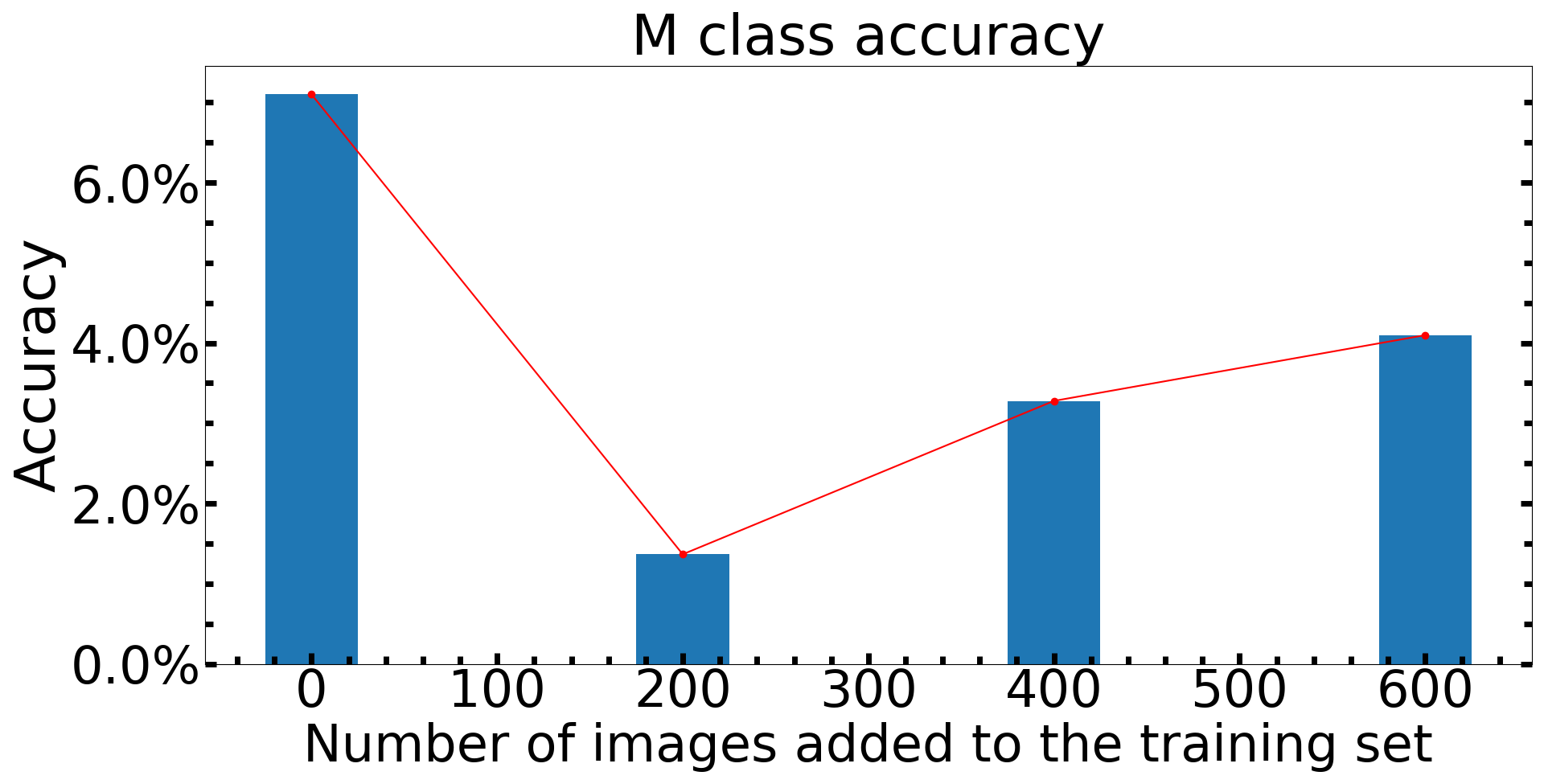}
        \caption{}
        \label{fig:accuracyvaraugM}
    \end{subfigure}
    \hfill 
    \begin{subfigure}{0.3\textwidth}
        \includegraphics[width=\linewidth]{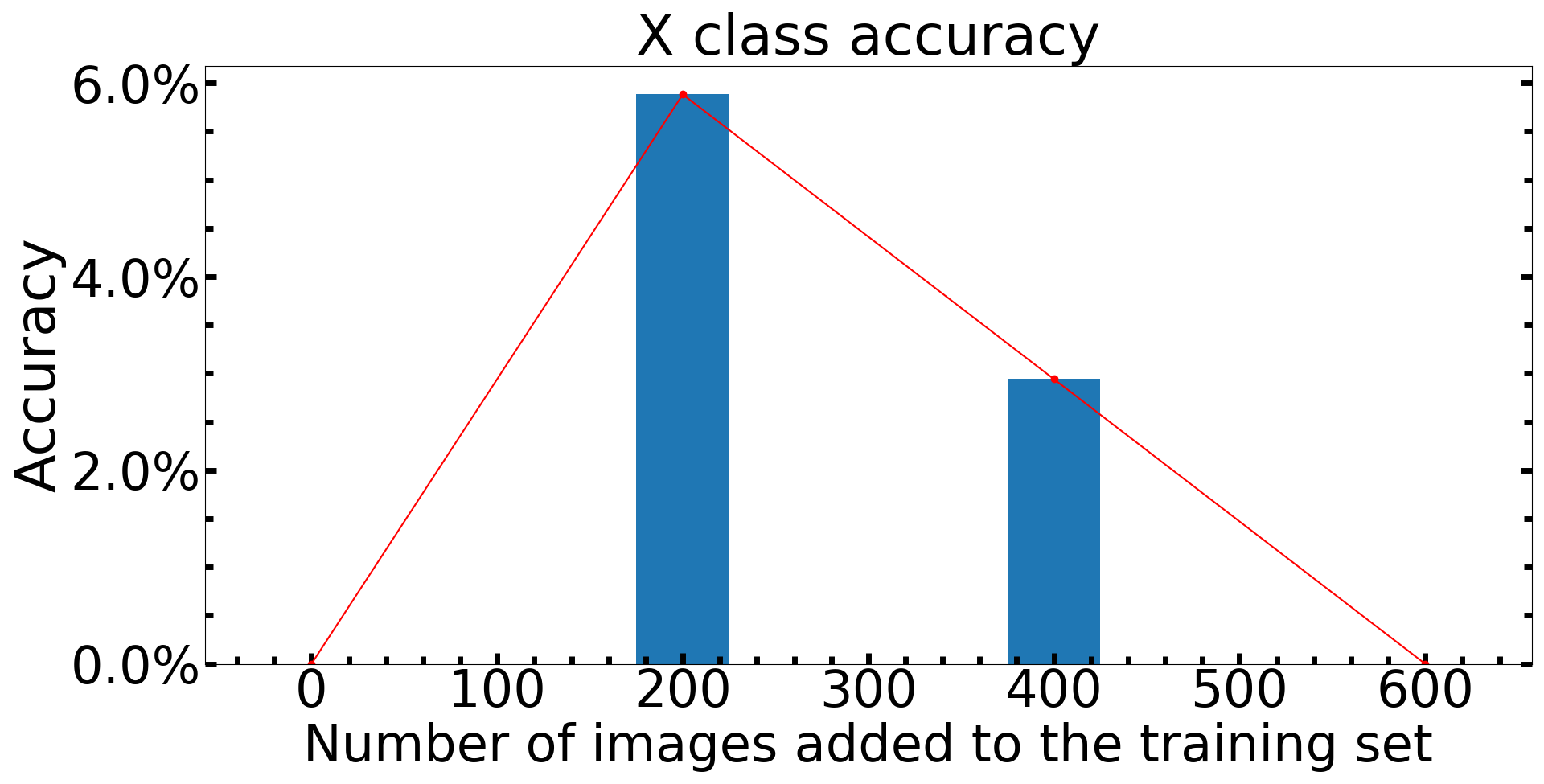}
        \caption{}
        \label{fig:classifier}
         \label{fig:accuracyvaraugX}
    \end{subfigure}
    \caption{Variation in accuracy per class, increasing the number of augmented samples added. Panel a) shows the evolution of the accuracy of A class, panel b) shows that of M class and panel c) shows that of X class.}
    \label{fig:accuracyvaraug}
\end{figure}

In Figure \ref{fig:accuracyvar}, we can see that adding the generated data to the training set increases the accuracy for the three least represented classes, A, M, and X. The soft grey regions are the variation in the accuracy, and for the A and M classes, the more images we add, the better the detection; for example, for A class with 600 images we reach 81.9\% accuracy compared to 30.1\% without adding any synthetic data, and for M class with 600 images we reach 29.1\% accuracy compared to 7.1\%. The is not exactly the same for the X class. Figure \ref{fig:accuracyvaraugX} appears to show that adding a lot of data leads to a decrease in the accuracy gained. At this stage, we are not making any assumptions as to how the detection accuracy is going to develop by adding more data of a particular class. The literature \citep{YANG2023100409} suggests that this depends on the initial size of the augmented classes, but also on the specific data used. We would need many further tests and more synthetic images to better explore this topic (e.g., adding 1000, 2000, 3000, or more images for all of the least represented classes).  

As an additional test, we contrasted the outcomes achieved by incorporating generated images with those obtained using classical data-augmentation techniques applied to authentic images, employing the same methodology of involving three incremental stages (200, 400, and 600). We employed various data-augmentation techniques through a series of transformation compositions using the torchvision library \citep{pytorch}. The techniques used include: random horizontal and vertical flipping, random rotation with varying degrees, random affine transformations with specified degrees, translations, and shearing. As shown in Figure \ref{fig:accuracyvaraug}, the application of classical augmented samples yields inconsistent results and does not consistently enhance performance. In certain instances, such as the M accuracy shown in \ref{fig:accuracyvaraugM}, there is a decline in performance compared to when no data-augmentation techniques are used.

\begin{figure}
\centering{%
\centering
 \includegraphics[width=\hsize]{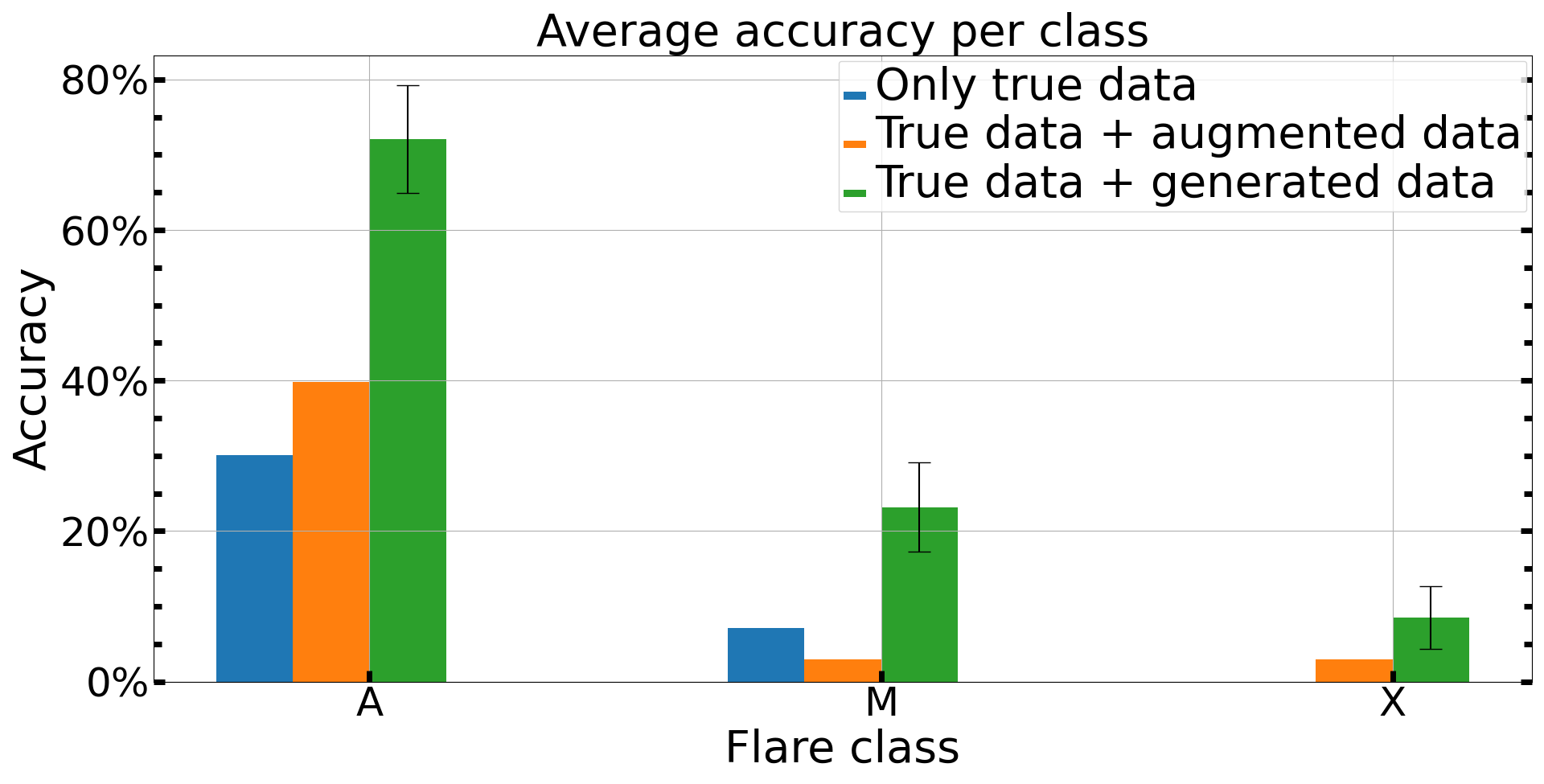}%
}
\caption{Accuracy values of the least represented class with the addition of synthetic images, with only true images and with data augmentation on true images in the training set.}
\label{fig:gen_nogen_compare_acc}
\end{figure} 

We are aware that using all these classical data transformations could result in significant changes to the data distribution, but we take all of them in order to be able to compare with the results obtained with the DDPM; otherwise using only vertical flipping or random rotation would not be enough for the lower represented class to produce 200, 400, 600 transformed images. In the case of the X class, our training set contains only 48 examples. This limitation makes it impossible to generate 200, 400, or 600 unique augmented data points without duplicating the same object, if we use only vertical flipping or random rotation. In the contrary using more transformations we avoid duplications but they lead to deviations from typical phenomena, leading to deteriorated results, as demonstrated in Figure \ref{fig:accuracyvaraug}. 

In contrast, augmentation using the diffusion model does not lead to this issue. With DDPM, we generate images that are not mere copies of the training data, as evidenced by the standard deviation maps in Figure \ref{fig:stdmap}. These maps show, for instance, that there are no flare phenomena occurring at the poles. As we can see in Figure \ref{fig:gen_nogen_compare_acc}, the addition of generated samples to the training set improves the performance of a supervised classifier by increasing its detection accuracy. Consequently, the technique utilised in this project is a valid method for overcoming the unbalanced dataset and for generating new images of the Sun in which we can control its level of activity.

\label{classfication_app}
\subsection{Solar flare prediction experiment}
Predicting solar flares is a critical task given the consequences outlined in Section \ref{sec:introduction}. Generally, it is posed as a classification problem (\citealp[]{Huang_2018, Li_2020, Pandey_2023}), where given input data \textbf{x} sampled at time $t_{0}$, the goal is to predict whether a flare will occur in the time window $t \in (t_{0}, t_{0} + \Delta t]$, with $\Delta t$ being arbitrarily chosen. There are various approaches to tackling this problem. For example, one approach is multi-class classification, the aim of which is to predict whether there will be an A, B, C, M, or X flare or their subclasses. Another approach is binary classification, where data are grouped based on the consequences of the solar flares; A, B, and C flares are grouped together, and M and X flares, which are more dangerous, are classified in another group. Further modelling criteria involve the use of either full-disc images \citep[as is done in these studies]{Pandey_2023, 2023ApJSYi}, or using patches to focus on the active regions \citep{Zheng_2019}. From a machine learning perspective, using patches of the active regions as input can potentially enhance model performance per active region due to their high resolution. However, from an 'artificial intelligence' standpoint, one would expect  the model to find out where to focus, eliminating the need for various preprocessing steps. Furthermore, it is possible to conduct a full-disc forecasts using a patch-based model and the output flare probabilities for each active region are typically aggregated. This approach treats all active regions independently and assigns them equal weight, which may not accurately reflect reality, as discussed in \citealp[]{pandey_2022, Pandey_2023}.

In this experiment, we conduct a full-disc solar flare prediction as a binary classification problem with a 24 hour time window. In this setup, A, B, and C flares are categorised as class 0, while M and X flares are categorised as class 1. Our objective with this experiment is not to achieve state-of-the-art efficiency in solar flare prediction, but rather to demonstrate the impact of using images generated by the DDPM, as illustrated in Section \ref{classfication_app}.

\begin{figure}
\centering
 \includegraphics[width=\hsize]{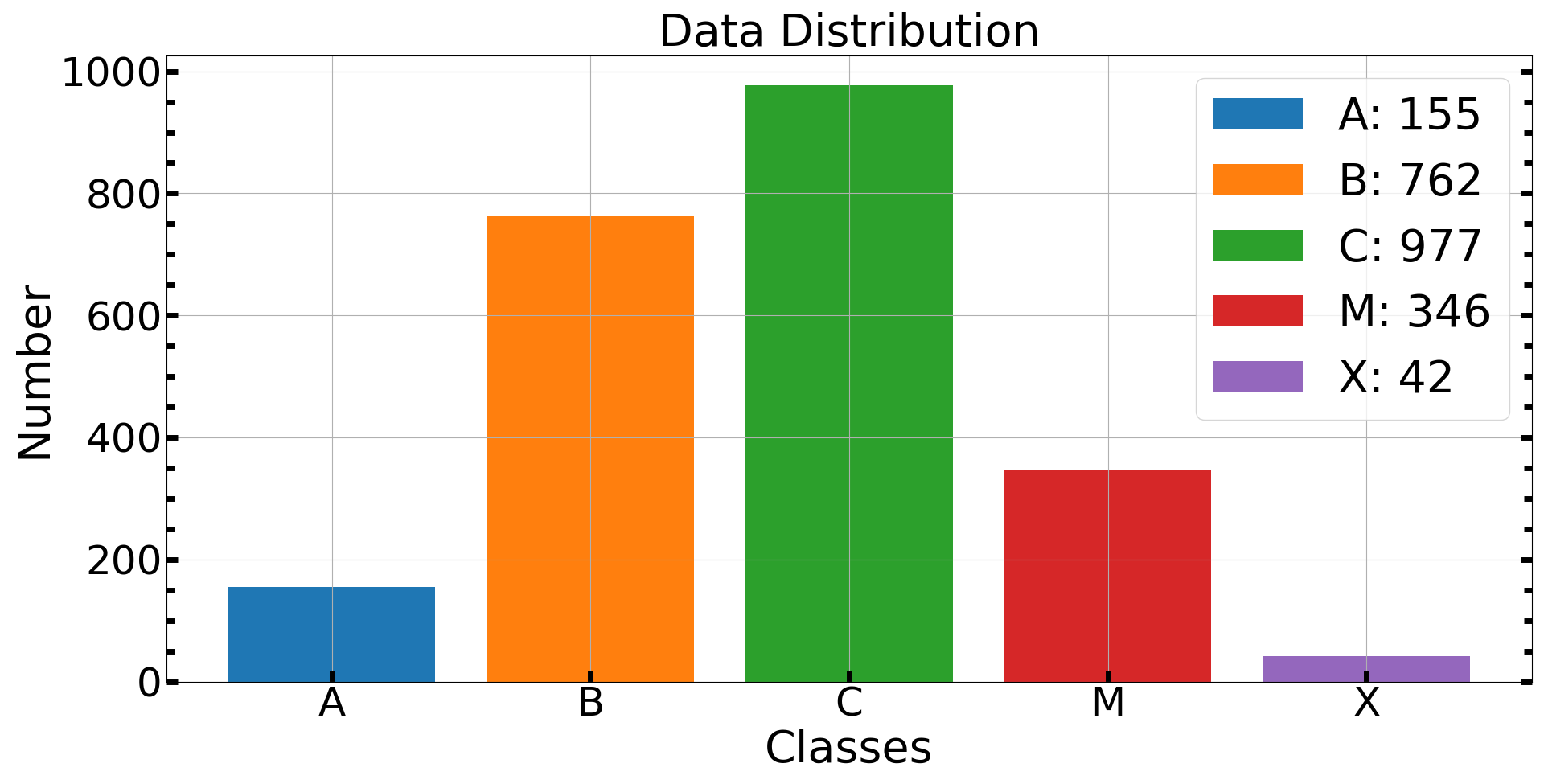}%
\caption{Histogram distribution of the labelled dataset with the discrete GOES labels: A, B, C, M and X.}
\label{fig:solar_flare_pred_dataset}
\end{figure}

However, the dataset described in Section \ref{sec:dataset} treats all flare events as independent samples -- even if they occur from the same flaring region. For flare prediction, we need to be more selective and select only one image per subsequent 24 hour window. For this reason we take a single image each day at 00:00:00 and label it as the most intensive flare that will occur in the next 24 h (e.g. if in the next 24h there is a C and an X flare then, we label the image as X). After this preprocessing, we end up with a total of 2282 data points that follow the distribution shown in Figure \ref{fig:solar_flare_pred_dataset}, which is the same  as in Figure \ref{fig:perc_goes_clas}. With this new dataset, we train a new conditional diffusion model with the same best setup found in Section \ref{sec:results} for generating the images. For the solar flare prediction architecture we keep the same DeIT backbone \citep{touvron2021training}, as described in Section \ref{sec:experiments}. This architecture is combined with a weighted Cross-Entropy loss (assigning 0.1 to the majority class and 0.9 to the minority class) and employs a learning rate that decays following a cosine function. We train the model for 18 epochs using the same initial setup as previously described. This training is conducted under three different scenarios: without augmented data, with classical data augmentations, and with DDPM data-augmentations. The classical data augmentation techniques used include only vertical flipping and random rotation within a range of 10 degrees. The DDPM augmentations involve injecting varying amounts of data into the training set, ranging from 50 to 500 instances for the under-represented classes (M and X), which are both classified as 1 in our binary classification scenario. Notably, this augmentation does not include additional data from the more prevalent A, B, and C classes. The metrics taken into consideration are:
\begin{itemize}
    \item the true skill statistics (TSS): \begin{equation}
        TSS = \frac{TP}{TP + FN} - \frac{FP}{FP + TN},
    \end{equation}
    \item the Heidke skill score (HSS): \begin{equation}
        HSS = 2 \times \frac{TP \times TN - FN \times FP}{(P \times (FN + TN) + (TP + FP) \times N)},
    \end{equation}  
    \begin{enumerate}
        \item N = TN + FP and P = TP + FN.
    \end{enumerate}
\end{itemize}
These metrics range from -1 to 1, where -1 indicates all incorrect predictions, 0 signifies performance equivalent to random guessing, and 1 denotes perfect predictions. Both metrics are employed in the context of solar-flare prediction because they are useful for assessing predictive performance, particularly in scenarios with imbalanced class distributions.

\begin{figure}
\centering
 \includegraphics[width=\hsize]{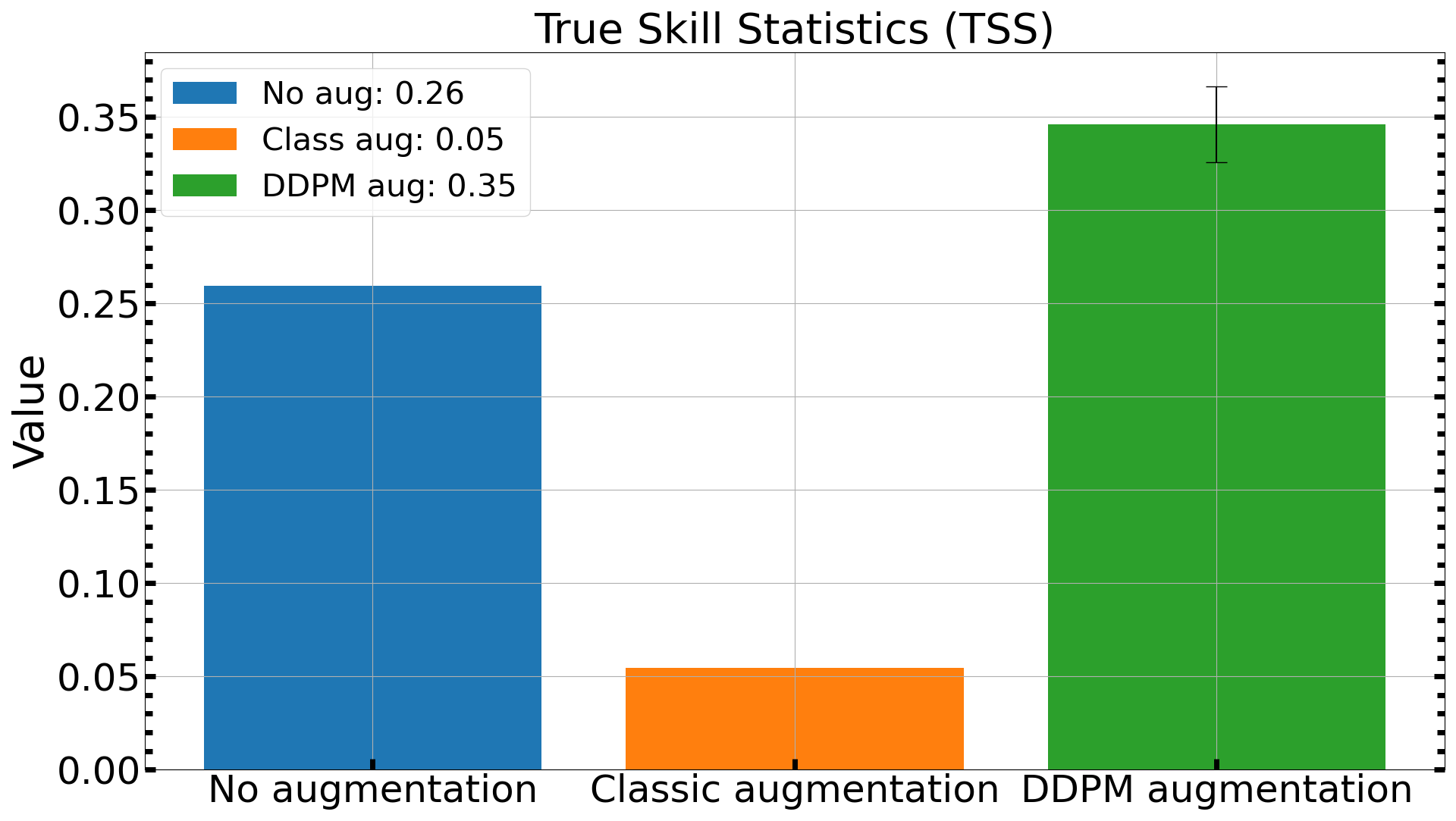}%
\caption{True skill statistics values in the three different scenarios: without data augmentation, with classical data augmentation and with DDPM data augmentation.}
\label{fig:tss}
\end{figure}
\begin{figure}
\centering
 \includegraphics[width=\hsize]{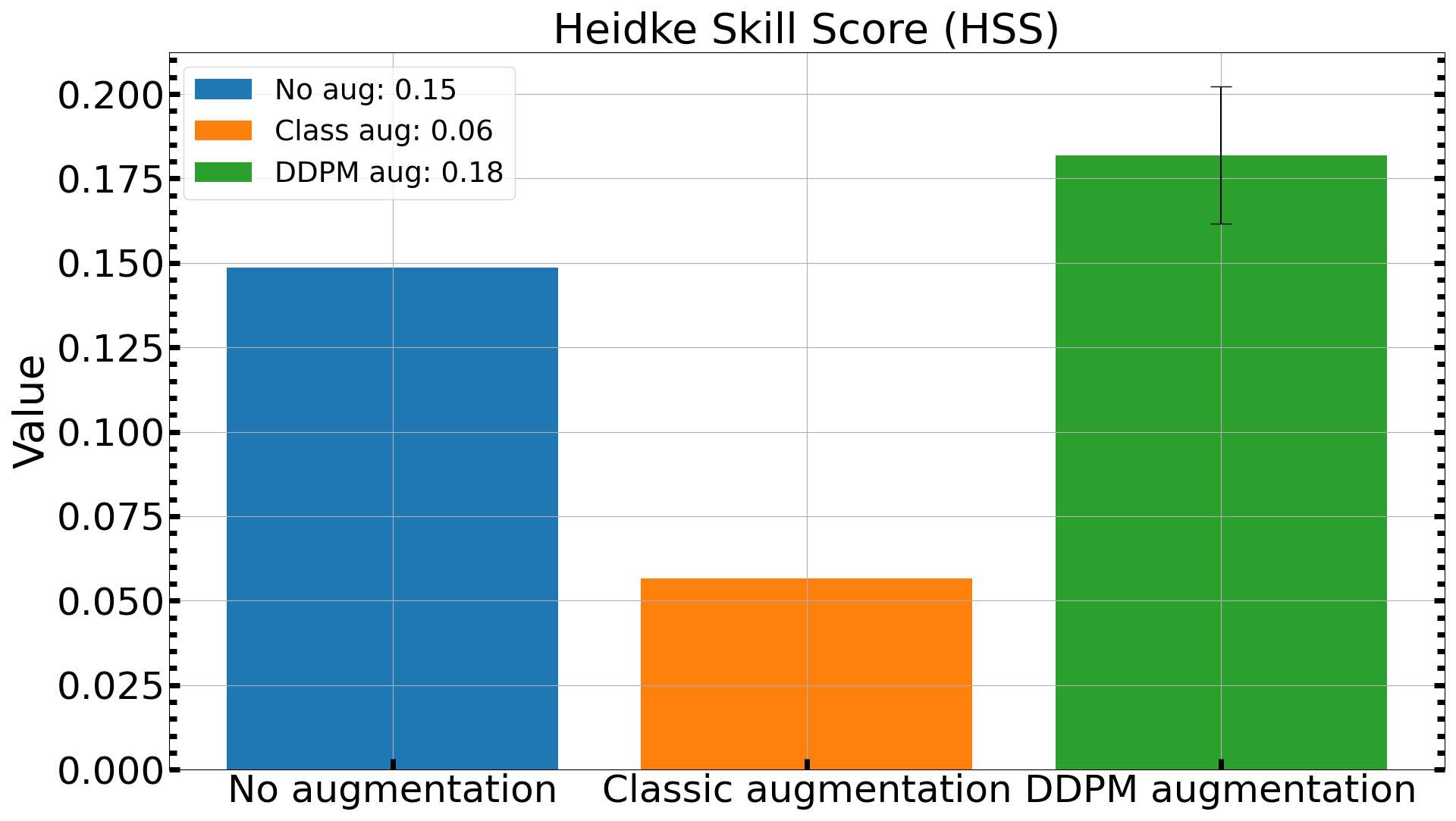}%
\caption{Heidke skill score values in the three different scenarios: without data augmentation, with classical data augmentation and with DDPM data augmentation.}
\label{fig:hss}
\end{figure} As observed in Figures \ref{fig:tss} and \ref{fig:hss}, employing DDPM augmentations consistently improves both the TSS and HSS metrics. The error bars represent the standard deviation calculated from using 50, 100, 200, 300, 400, and 500 generated data points per least-represented class in the training set, while the bars indicate the mean values. The final values, with DDPM-augmented data, are 0.35 $\pm$ 0.02 for the TSS and 0.18 $\pm$ 0.02 for the HSS. Classical data augmentation, instead, consistently decreases the performance, even when using data-augmentation techniques that should not deviate significantly from the normal distribution of the flares, such as vertical flipping and random rotation within 10 degrees. All the evaluation metrics are computed on the same test set, where there are only true data and no augmentations of any kind.

\section{Conclusions}
\label{sec:conclusions}
The goal of this work is to show the ability of the DDPM to generate images conditioned on the flare class so that they can be used in an equivalent way to the true images and thus prevent dataset imbalance towards the highest energy flares. It is possible to see from  Figure \ref{fig:latentspace} -- where different architectures are used to encode the image information from the 64x64 images to compute the metrics -- that the ceVAE architecture presents some clustering and differentiation between the different flare classes with respect to the other architecture, and thus it is possible to highlight some differences between various classes even with a 64x64 image. Undoubtedly, in Figure \ref{fig:latentspace}, the ceVAE latent space is not perfectly clustered and in a future work we will analyse the effects of increasing the image size and whether or not this will lead to a more definite clustering. The results are presented in in Table \ref{results_metrics}.

In this Table, we trained a classifier on authentic data and subsequently evaluated on generated data. We find that the model successfully generates X-flare Sun instances that are very similar to authentic ones, despite the image dimensions being limited to 64x64. This suggests that the model effectively recognises the distinctions between various flare classes despite the limited image size. It is noteworthy that the average time interval between successive images in our dataset is 72 minutes. Consequently, certain images may exhibit minimal visual differences while being associated to distinct flare classes. The DDPM does not encounter any issues in this scenario, as it does not involve classification. Whenever this model is used, the related flare class is always provided as input with the image. Therefore, even if two images appear visually similar, the presence of the flare class serves as a discriminant. On the contrary, the application will not be able to distinguish all the data correctly, leading to greater uncertainty (Table \ref{results_metrics}). This uncertainty will be larger if we consider images that come from the same active region. The primary objective of this application is to demonstrate that by solely training a classifier without any fine tuning of the model, we were able to enhance performance in terms of the metrics employed here by using the synthetic images to balance the dataset. For this reason, we decided to subset the dataset in such a way that it is standardised for solar-flare prediction and to test the DPPM in this scenario. As is true for the classification task, in the solar-flare prediction task the use of the DDPM-augmented data improves the performance of the model when using the same setup as without data augmentation.

In future work, we would like to better comprehend the generation capabilities of the DDPM models (e.g. analysing the DDPM latent space), apply them to image-to-image translation tasks \citep[e.g. to obtain HMI magnetograms from each generated image]{saharia2022palette}, and to increase the image size to explore the impacts of this change. In addition, we would like to overcome the dataset limitations described in Section \ref{sec:dataset} and zoom in on the flaring regions, validating them with physical metrics so that they can be used for physics and machine learning-related downstream tasks.  \citep{Armstrong2019, 10.3389/fspas.2020.00034, angeo-39-861-2021}.

\begin{acknowledgements}
      This research was partially funded by the SNF Sinergia project (CRSII5-193716): Robust Deep Density Models for High-Energy Particle Physics and Solar Flare Analysis (RODEM).
\end{acknowledgements}

\bibliographystyle{aa} 
\bibliography{references} 

\newpage
\begin{appendix} 

\section{Dataset limitation}
\label{data_limitation}
\begin{figure*}
    \begin{minipage}{0.5\textwidth}
        \centering
        \begin{subfigure}{\linewidth}
            \includegraphics[width=\linewidth]{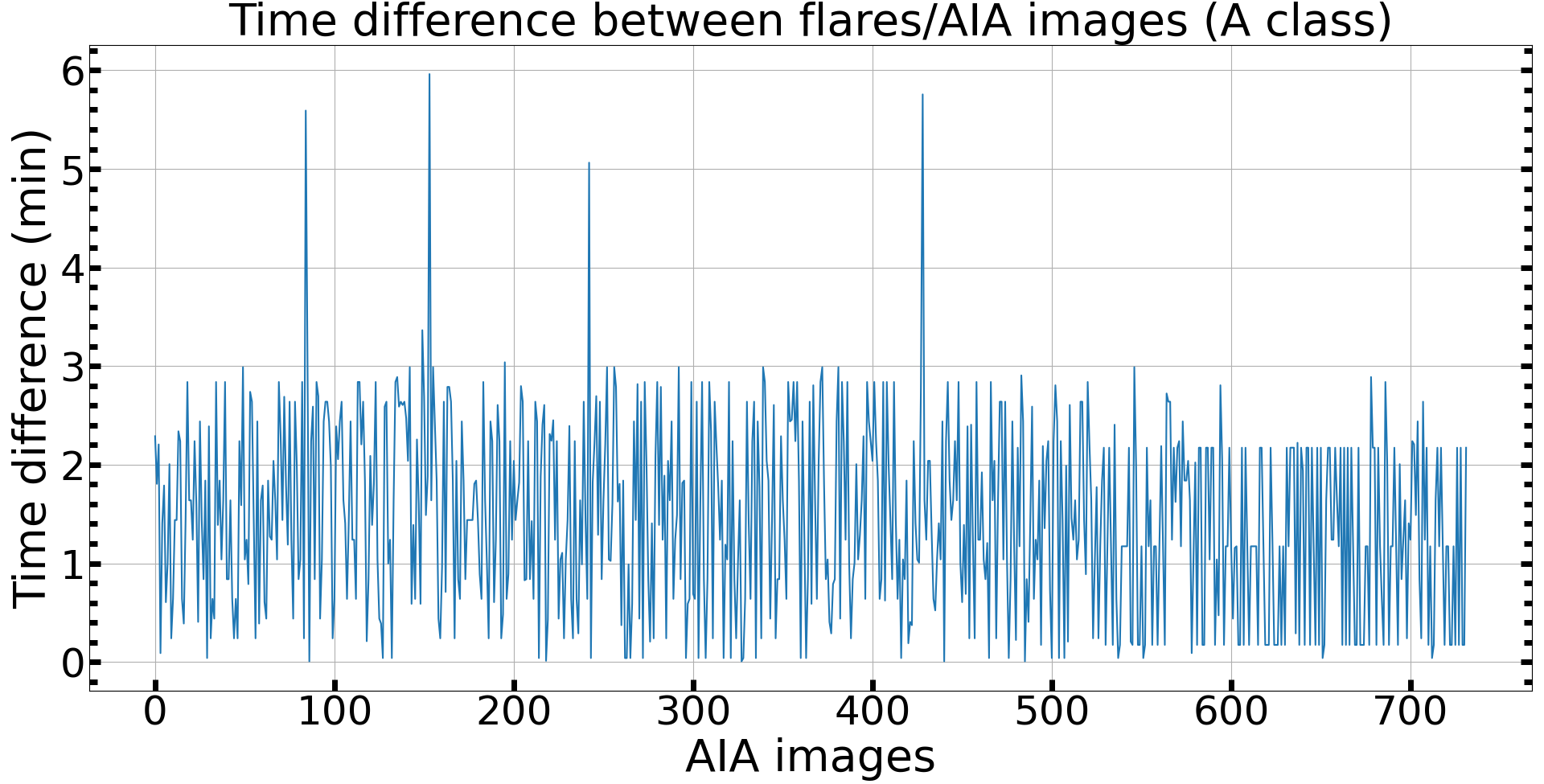}
            \caption{}
            \label{fig:clipA}
        \end{subfigure}
    \end{minipage}
    \begin{minipage}{0.5\textwidth}
        \centering
        \begin{subfigure}{\linewidth}
            \includegraphics[width=\linewidth]{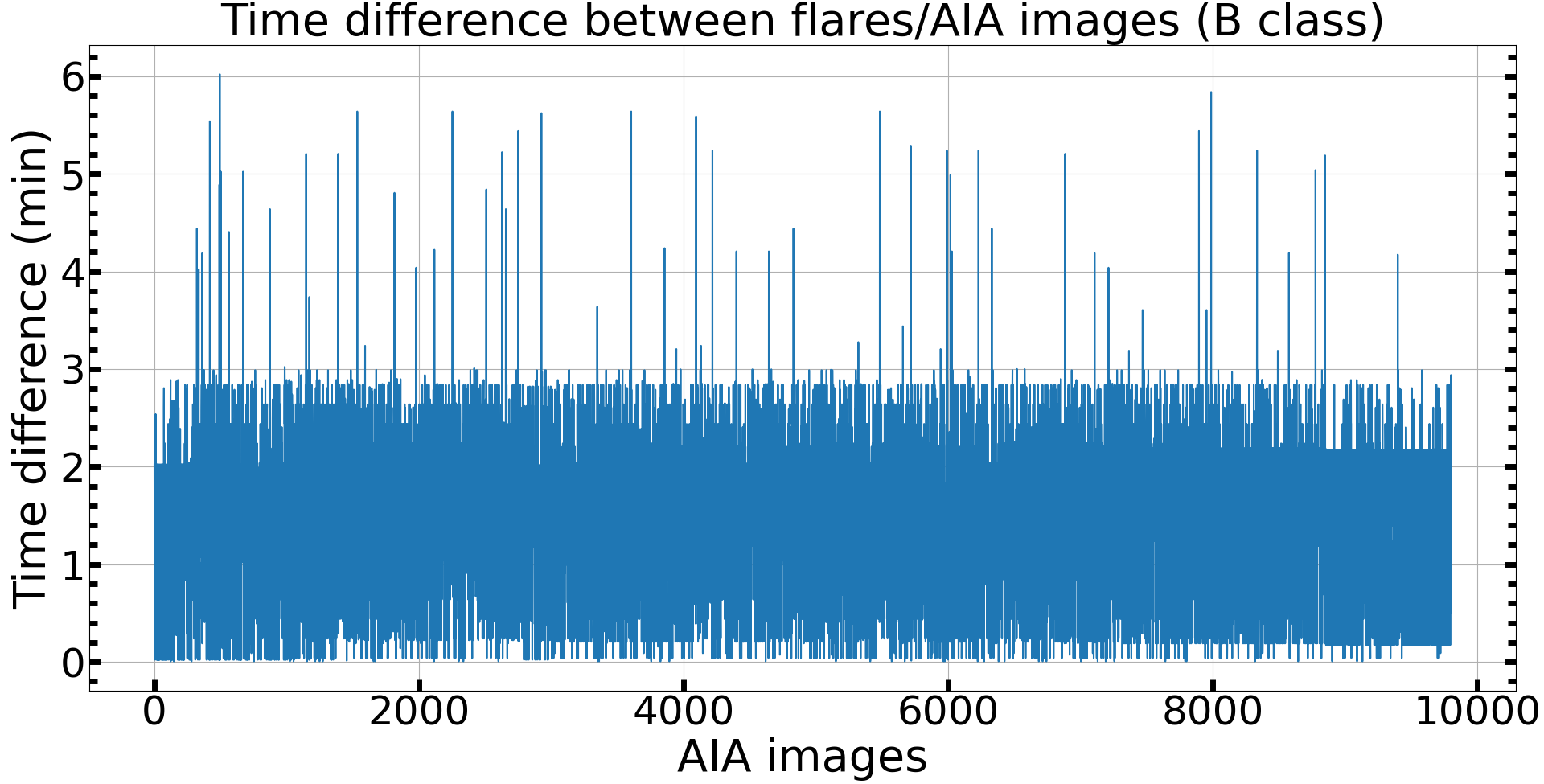}
            \caption{}
            \label{fig:ceVaeB}
        \end{subfigure}
    \end{minipage}
    
    \vspace{\baselineskip} 
    \begin{center}

    \begin{minipage}{0.5\textwidth}
        \centering
        \begin{adjustbox}{valign=M,center} 
            \begin{subfigure}{\linewidth}
                \includegraphics[width=\linewidth]{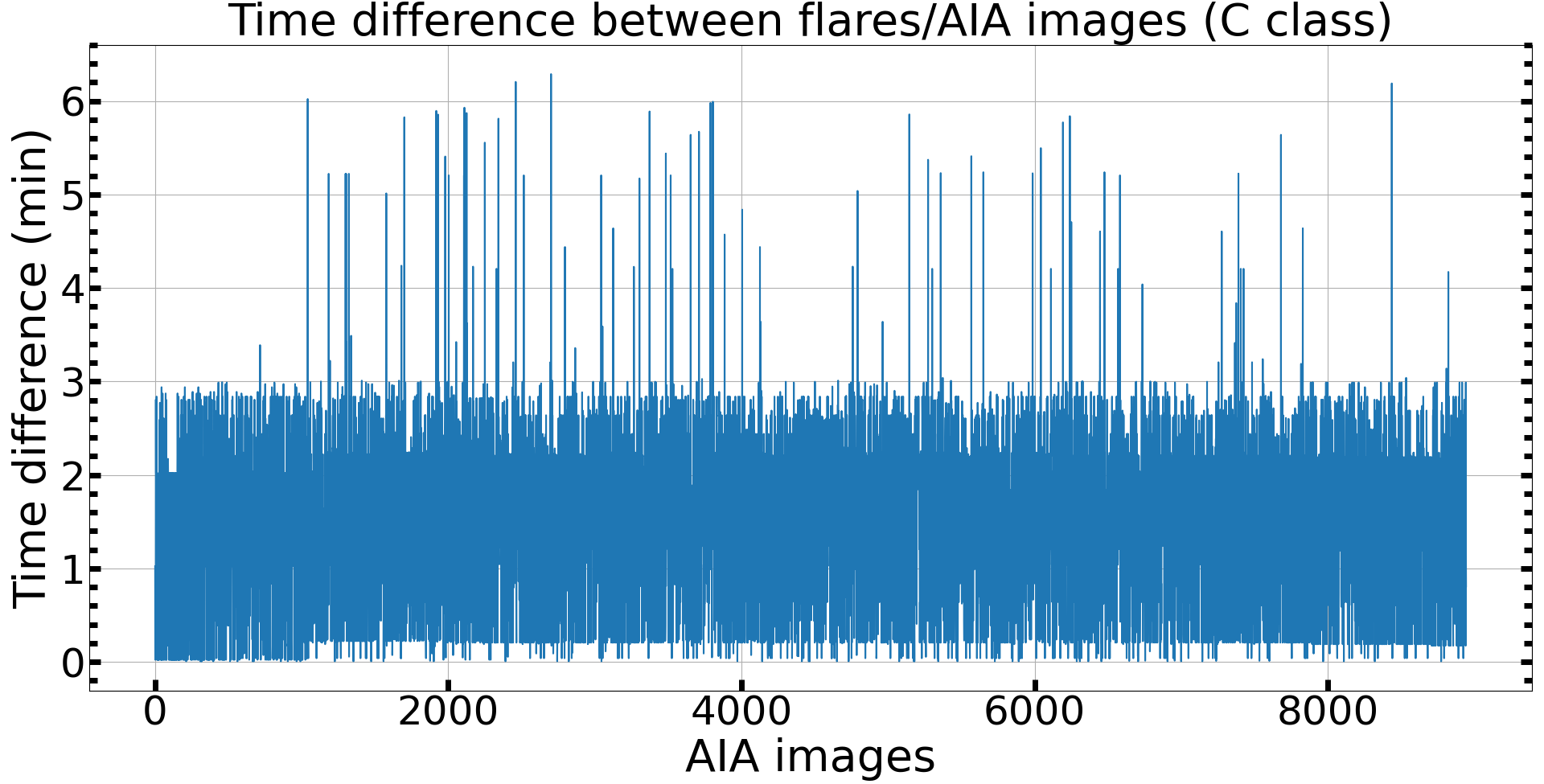}
                \caption{}
                \label{fig:classifierC}
            \end{subfigure}
        \end{adjustbox}
    \end{minipage}
    \end{center}
     
    \vspace{\baselineskip} 
    
    \begin{minipage}{0.5\textwidth}
        \centering
        \begin{subfigure}{\linewidth}
            \includegraphics[width=\linewidth]{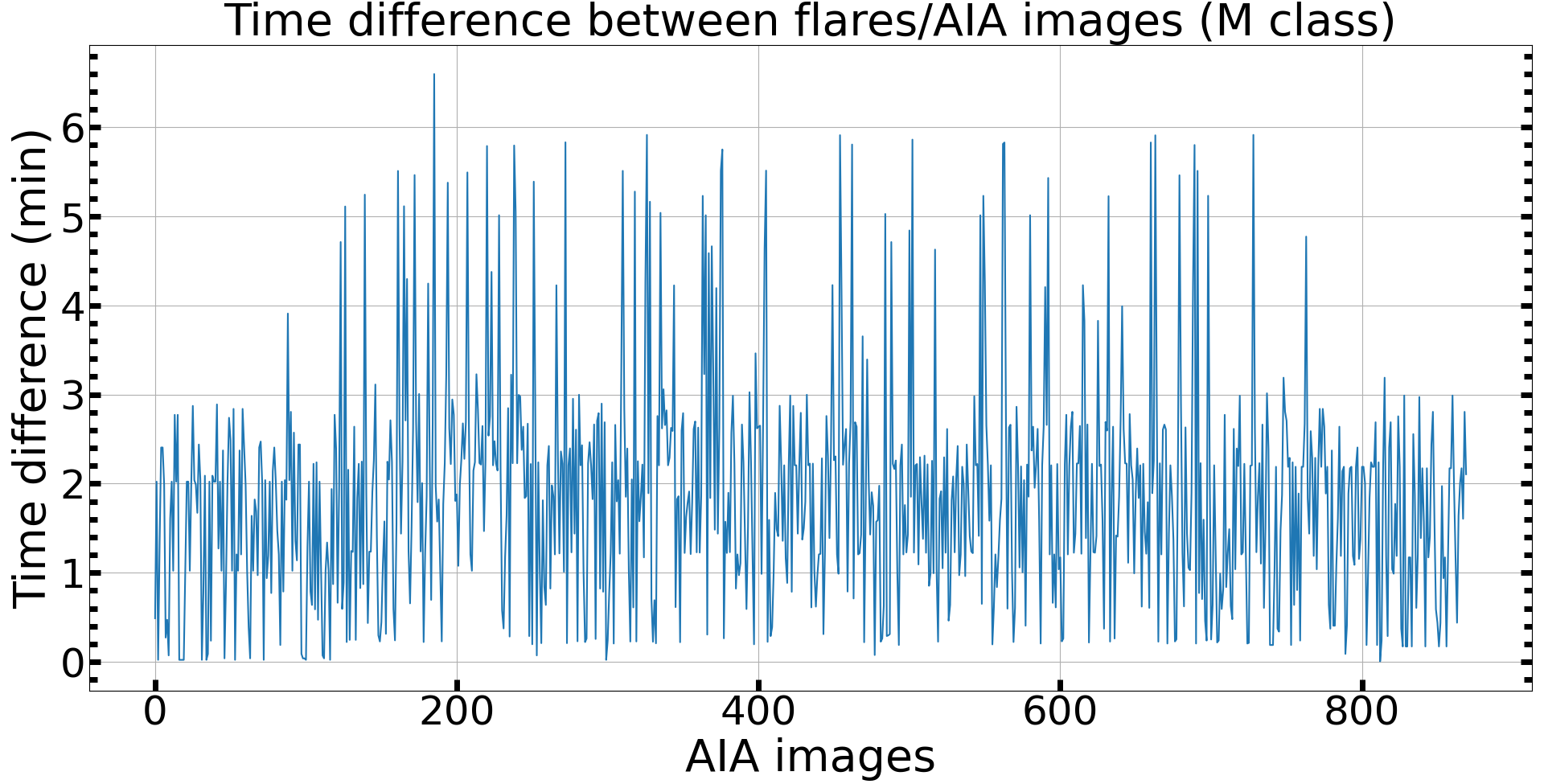}
            \caption{}
            \label{fig:classifierM}
        \end{subfigure}
    \end{minipage}
    \begin{minipage}{0.5\textwidth}
        \centering
        \begin{subfigure}{\linewidth}
            \includegraphics[width=\linewidth]{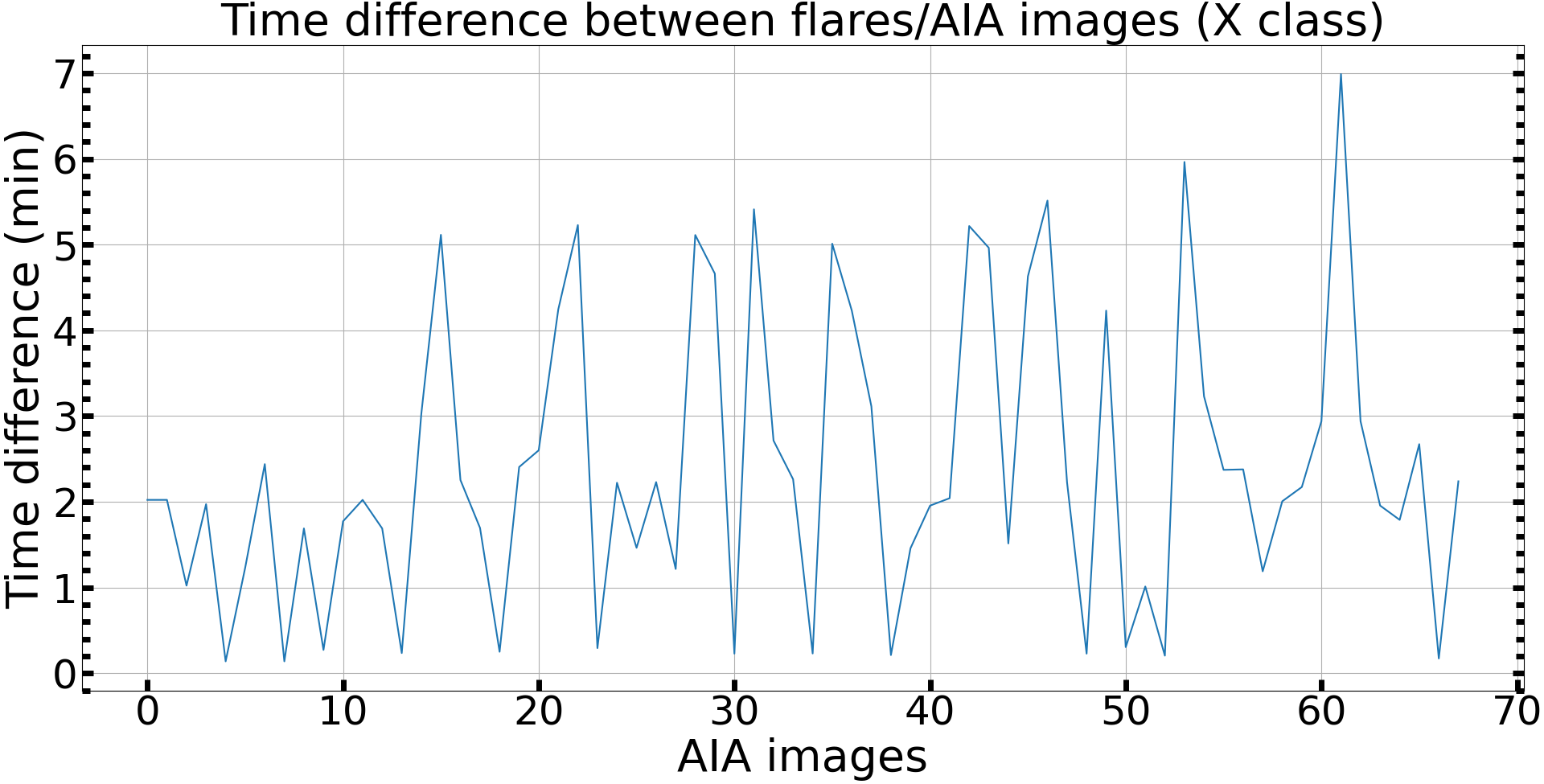}
            \caption{}
            \label{fig:classifierX}
        \end{subfigure}
    \end{minipage}
    
    \caption{Time-delay histograms per flare class of the AIA image with respect to the peak time of the flaring event. Panels a) to e) represent the time delay of the images belonging to classes A to X, respectively.}
    \label{fig:time_delay}
\end{figure*}

As introduced in Sect. \ref{sec:data_sel}, the limitation of our dataset is related to the time resolution of the AIA instrument in the SDOMLv2, which is 6 minutes instead of 12 seconds as in the original SDO data. As mentioned by \citealp{Galvez_2019}, this is done in order to perform the temporal synchronisation with the EVE instrument. This is a limitation due to the fact that we are interested in a specific time when searching for the SDO image, as we only want to consider flaring occurrences. In fact, when cross-correlating the HEK dataset with the SDOMLv2 dataset, we use a time tolerance of 7 minutes to maximise the number of images, while bearing in mind that we may lose flaring information if the closest image is more than 10 minutes away in time. In Figure \ref{fig:time_delay}, the time delays per class are depicted.

\section{Architecture}
\label{model_arch}

\begin{figure}[]
\centering
 \includegraphics[width=\hsize]{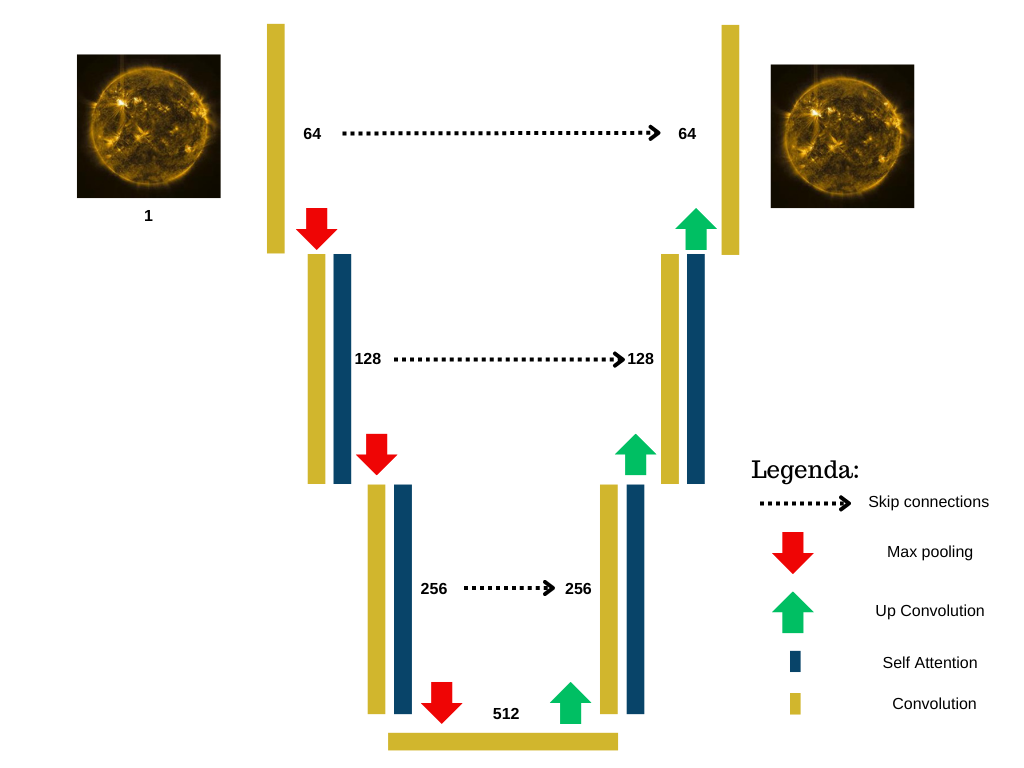}%
\caption{Unet architecture implementation.}
\label{fig:unet}
\end{figure}

The U-Net architecture is a form of convolutional neural network (CNN) that was initially developed for biomedical-image segmentation, but has since been applied to other image-segmentation issues as well. The network is known as U-Net because its architecture is U-shaped, with a contracting path (encoder) on the left and an expanding path (decoder) on the right. The contracting path is comprised of convolutional and pooling layers that gradually decrease the spatial resolution of the input image, whereas the expanding path employs upsampling and convolutional layers to gradually increase the resolution and generate a segmentation mask. In addition, U-Net includes skip connections that directly link the layers between the encoder and decoder channels. These skip connections enable the network to propagate information from the contracting path to the expanding path at varying spatial resolutions, thereby preserving high-resolution characteristics. In conclusion, our implementation (Figure \ref{fig:unet}) between every downsampling and upsampling layer, includes a self-attention layer \citep{vaswani2017attention}, which is used to model long-range dependencies between different spatial locations in an image. In this instance, the self-attention mechanism computes the relative importance of each spatial location in an image relative to other spatial locations. This is accomplished by applying a set of learned weight vectors to the input feature map to generate a set of attention maps that indicate the importance of each spatial location. The attention maps are then used to re-weight the input feature map, emphasising the most significant spatial locations and omitting the less significant ones. This generates a new feature map that contains the most pertinent data for the image-generation assignment.

\section{Image generation with 128x128 pixel resolution}
\label{128x128}
\begin{figure*}
\centering
\subfloat[\label{fig:clip}]{%
 \includegraphics[width=0.3\hsize]{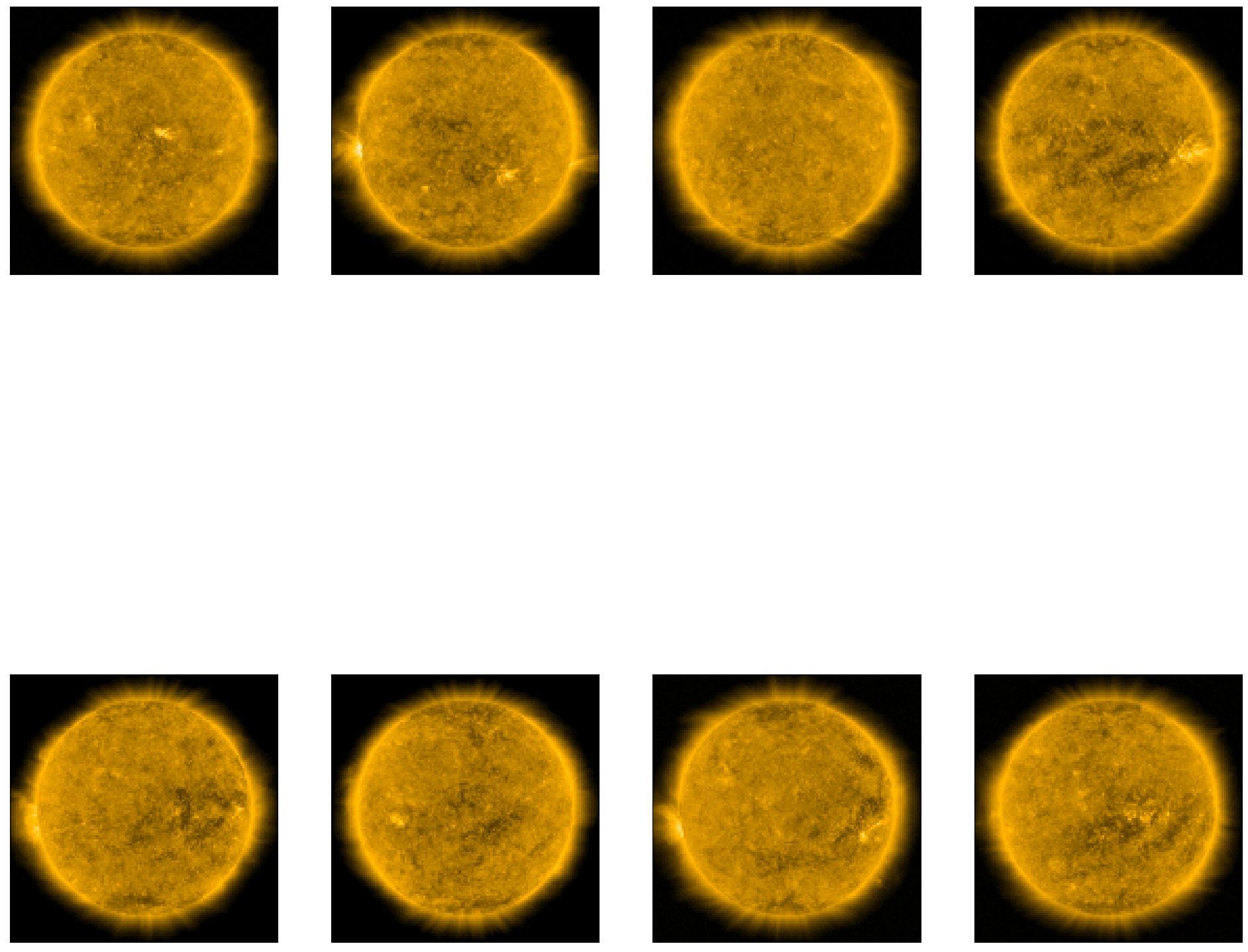}%
}
\qquad
\subfloat[\label{fig:ceVae}]{%
 \includegraphics[width=0.3\hsize]{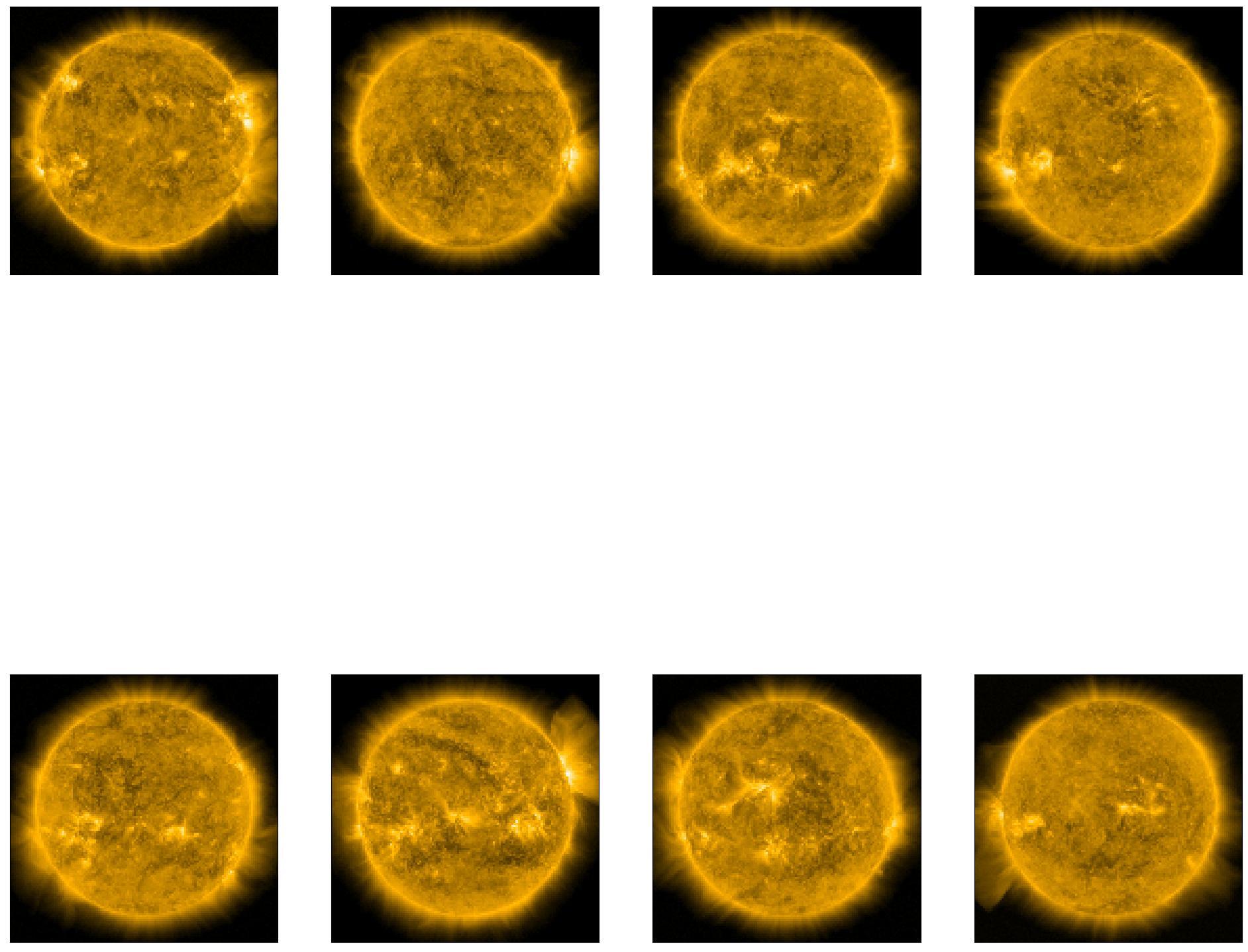}%
}
\qquad
\subfloat[\label{fig:classifier}]{%
 \includegraphics[width=0.3\hsize]{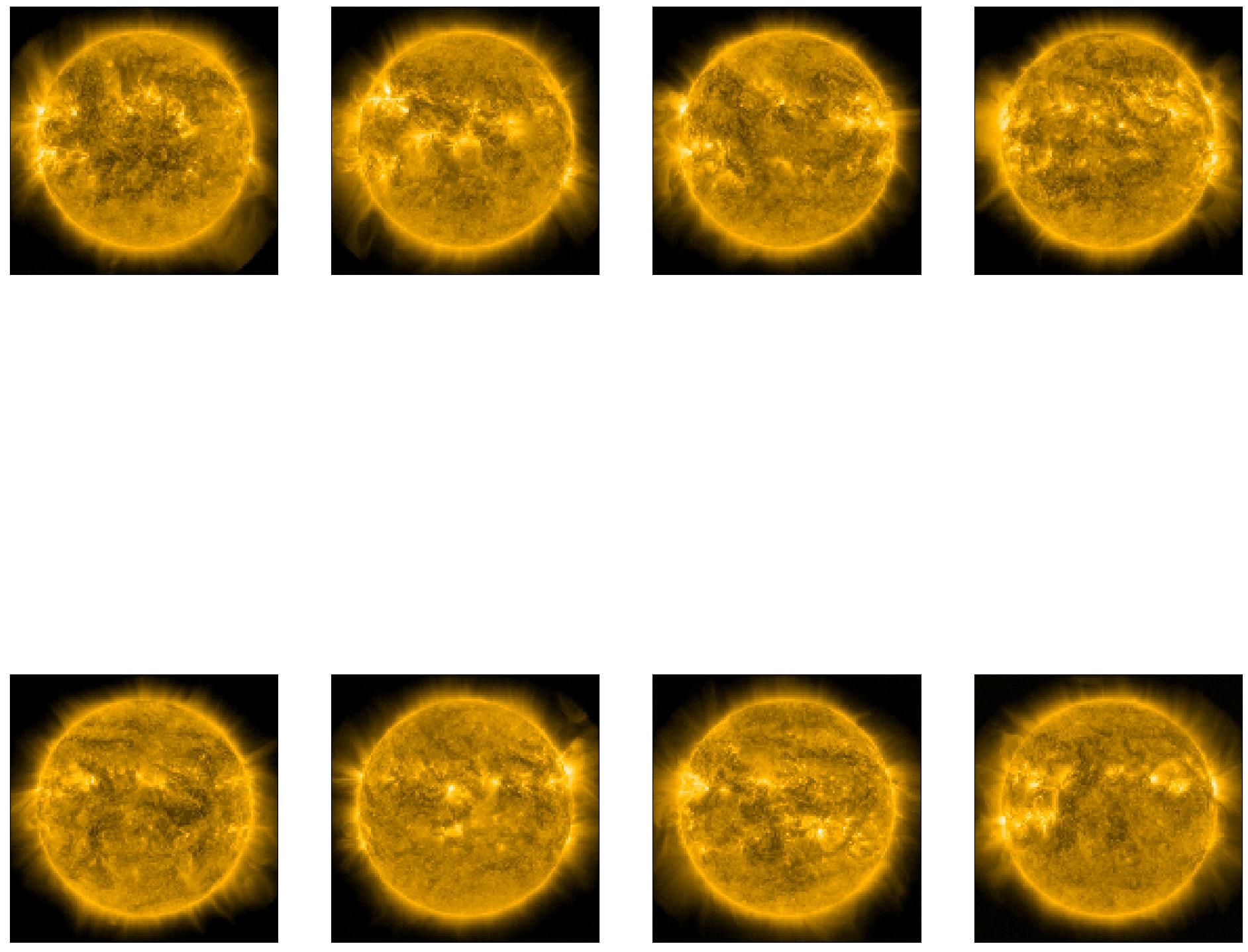}%
}
\caption{Generated images with 128x218 pixel resolution. Low level refers to A-class flares, medium level to B-class flares and high level to C-, M- and X-class flares. Panel a) shows the low-level activity, b) medium-level activity and c) the high-level activity.}
\label{fig:128x128_img}
\end{figure*}

Denoising diffusion probabilistic models are very computationally expensive, but are very good in manipulating the details \citep{dhariwal2021diffusion}. Indeed, increasing the resolution from 64x64 to 128x128, we can see (Figure \ref{fig:128x128_img}) that the generated images do not introduce physical artefacts at first sight, but further analysis should be carried in this regard. On the other hand, to be able to perform this generation, we cannot use the NVIDIA TITAN X GPU due to vram shortage (12 GB), but we use the NVIDIA A100 GPU with a vram of 40 GB. Despite this, we decrease the complexity of our architecture, removing two self-attention layers in the U-Net.

\section{F1-score, precision and recall}
\label{metricsperclass}
\begin{table}[h]
\centering
\begin{tabular}{|c|c|c|c|}
\hline
Class & F1-score & Precision & Recall \\
\hline
A & 0.80 & 0.76 & 0.83 \\
\hline
B & 0.26 & 0.50 & 0.17 \\
\hline
C & 0.55 & 0.54 & 0.56 \\
\hline
M & 0.70 & 0.74 & 0.65 \\
\hline
X & 0.42 & 0.35 & 0.51 \\
\hline
\end{tabular}
\caption{Metric results of classifier trained and tested on true data.}
\end{table}

\begin{table}[h]
\centering
\begin{tabular}{|c|c|c|c|}
\hline
Class & F1-score & Precision & Recall \\
\hline
A & 0.77 & 0.70 & 0.85 \\
\hline
B & 0.24 & 0.40 & 0.17 \\
\hline
C & 0.28 & 0.23 & 0.35 \\
\hline
M & 0.37 & 0.30 & 0.47 \\
\hline
X & 0.26 & 0.46 & 0.18 \\
\hline
\end{tabular}
\caption{Metric results of the classifier trained and tested on generated data from the model with discrete GOES labels.}
\end{table}

\begin{table}[h]
\centering
\begin{tabular}{|c|c|c|c|}
\hline
Class & F1-score & Precision & Recall \\
\hline
A & 0.67 & 0.58  & 0.79 \\
\hline
B & 0.23 & 0.34 & 0.18 \\
\hline
C & 0.32 & 0.27 & 0.39 \\
\hline
M & 0.45 & 0.38 & 0.56 \\
\hline
X & 0.04 & 0.13 & 0.02 \\
\hline
\end{tabular}
\caption{Metric results of the classifier trained and tested on generated data from the model with discrete GOES labels and the ceVAE embeddings.}
\end{table}

\begin{table}[h]
\centering
\begin{tabular}{|c|c|c|c|}
\hline
Class & F1-score & Precision & Recall \\
\hline
A & 0.19 & 0.21 & 0.17 \\
\hline
B & 0.16 & 0.19 & 0.14 \\
\hline
C & 0.24 & 0.20 & 0.32 \\
\hline
M & 0.23 & 0.19 & 0.27 \\
\hline
X & 0.12 & 0.19 & 0.09 \\
\hline
\end{tabular}
\caption{Metric results of the classifier trained and tested on generated data from the model with the xray values.}
\end{table}
\end{appendix} 

\end{document}